\documentclass[10pt,a4paper]{article}
\usepackage[utf8]{inputenc}
\usepackage{amsmath,amssymb}
\usepackage{graphicx,psfrag,color}
\usepackage{amsfonts}
\usepackage{authblk}
\usepackage{pstricks,pst-node,pst-text,pst-3d}
\usepackage{caption}
\usepackage{subfigure}
\usepackage{setspace}
\usepackage[section]{placeins}
\usepackage{ytableau}
\usepackage{epstopdf}
\usepackage{graphicx}
\usepackage{lscape}
\usepackage{tikz}
\usepackage{physics}
\usepackage[normalem]{ulem}
\usepackage{hyperref}
\hypersetup{
	colorlinks=true,
	linkcolor=blue,
	citecolor=blue,
	filecolor=black,
	urlcolor=black,
}
\usepackage{enumerate}
\usepackage[left=2.5cm, top=3cm, right=2cm, bottom=2cm]{geometry}
\usepackage{cite}

\renewcommand\footnotemark{}

\renewcommand{\baselinestretch}{1.0}
\newcommand{\im}{\mathrm i}

\newcommand{\diag}{\operatorname{diag}}

\newcommand{\eq}{\begin{equation}}
\newcommand{\en}{\end{equation}}
\newcommand{\bear}{\begin{eqnarray}}
\newcommand{\ear}{\end{eqnarray}}

\newcommand{\sh}{\mathrm{sh}}

\newcommand{\ee}{\mathrm{e}}
\newcommand{\bt}{\begin{ytableau}}
\newcommand{\et}{\end{ytableau}}

\newcommand{\lp}{\left(}
\newcommand{\rp}{\right)}
\newcommand{\sft}{\delta}
\newcommand{\sg}{s}
\newcommand{\ind}{\iota}
\newcommand{\unknown}{{EAF}}

\makeatletter
\newcommand{\vast}{\bBigg@{3}}
\newcommand{\vastt}{\bBigg@{4}}
\newcommand{\Vast}{\bBigg@{5}}
\newcommand{\Vastt}{\bBigg@{6}}
\newcommand{\Vasttt}{\bBigg@{7}}
\newcommand{\Enor}{\bBigg@{8}}
\newcommand{\Enorr}{\bBigg@{9}}
\makeatother

\makeatletter
\renewcommand*\env@matrix[1][\arraystretch]{%
	\edef\arraystretch{#1}%
	\hskip -\arraycolsep
	\let\@ifnextchar\new@ifnextchar
	\array{*\c@MaxMatrixCols c}}
\makeatother

\allowdisplaybreaks[1]
\numberwithin{equation}{section}

\title{\mbox{}Canonical formulation for the thermodynamics of $sl_n$-invariant integrable spin chains}
\author[1,2]{T. S. Tavares\thanks{E-mail: tavares@df.ufscar.br}}
\author[1]{I. R. Passos\thanks{E-mail: ingrydrpassos@gmail.com}}
\author[1]{A. Kl\"{u}mper\thanks{E-mail: kluemper@uni-wuppertal.de}}
\affil[1]{\small Fakult\"{a}t f\"{u}r Mathematik und Naturwissenschaften, Bergische Universit\"{a}t Wuppertal, 42119 Wuppertal, Germany}
\affil[2]{\small Departamento de F\'{i}sica, Universidade Federal de S\~ao Carlos, 13565-905 S\~ao Carlos-SP, Brazil}
\date{}

\begin{document}
\renewcommand{\baselinestretch}{1.2}
	\maketitle
	\thispagestyle{empty}

\begin{abstract}
Integrable quantum spin chains display distinctive physical properties
 making them a laboratory to test and assess different states of
matter. The study of the finite temperature properties is
  possible by use of the thermodynamic Bethe ansatz, however at the expense of
dealing with non-linear integral equations for, in general, infinitely many
auxiliary functions.  The definition of an alternative finite set of auxiliary functions
allowing for the complete description of their thermodynamic properties at
finite temperature and fields has been elusive. Indeed, in the context of
$sl_n$-invariant models satisfactory auxiliary functions have been established
only for $n \leq 4$. In this paper we take a step further by proposing a
systematic approach to generate finite sets of auxiliary functions for
$sl_n$-invariant models. We refer to this construction as the canonical
formulation. The numerical efficiency is illustrated for $n=5$, for which we
present some of the thermodynamic properties of the corresponding spin chain.
\end{abstract}

\newpage

\section{Introduction}
 Almost a century has now passed since Bethe’s seminal work in
which he first transformed the solution of the one-dimensional Heisenberg spin
chain into a system of nonlinear algebraic equations \cite{BETHE}. The
understanding of this model and of other similar systems has developed
since. Lieb identified the same eigenvectors of the Hamiltonian as belonging
to a transfer matrix of a classical statistical model \cite{LIEB}, and not
long after Baxter put forward the idea of commuting transfer matrices
\cite{BAXTER} which naturally become generators of conserved charges. By now,
Yang-Baxter integrable spin chains and other exactly solvable models form a
field of their own. The question remains to what extent quantum integrability
is an asset in determining physical and mathematical properties.

In this respect, the treatment of fully interacting integrable systems is a
much more intricate task than for the case of free fermion
models. Integrability imposes factorization of the multi-particle $S$-matrix
into 2-particle $S$-matrices, however with non-trivial scattering. To mention
from the many physically relevant properties a most important example,
consider the evaluation of correlation functions for arbitrary distances,
fields and temperature which is, even from a wide point of view, successful  so far only for a few
integrable systems \cite{GOHMANN,BABENKO}.  Each of the most prominent
integrable models inspired the development of a range of different (and
sometimes model-specific) techniques, which is very much exemplified by the
several formulations of Bethe ansatz that emerged so far
\cite{LENINGRADO,BAXTERB,RESHE,KOREPINB,MARTINS,CAO,MAILLET}.  On one hand it
evinces ingenuity in the field, on the other it demonstrates a remaining lack
of a universal comprehension.

 Similarly, the evaluation of thermodynamical properties, like
other physical properties, has not escaped this situation. In
  principle, once a model is identified as Yang-Baxter integrable and a Bethe
  ansatz is established, all what is left is to solve the Bethe equations in
  order to obtain the eigenvalues for evaluating the partition function.
  This is the basis of the so-called Thermodynamical Bethe Ansatz (TBA)
  \cite{YANGTe,GAUDIN,TAKAHASHI}. Such a program leads to the study of all
  patterns of Bethe roots corresponding to bound states, which culminates into
  the string-hypothesis for integrable quantum chains \cite{TAKAHASHI}. An
  alternative procedure, known as the Quantum Transfer Matrix (QTM) method,
  allows for the evaluation of the thermodynamical potential through a single
  eigenvalue of a new object \cite{TROTTER,MSUZUKI,KLUMPER92,KLUMPER93}.

Among the criticisms to the former method are the lack of proof of the string
hypothesis and the fact that one needs to solve an infinite number of
nonlinear integral equations (NLIEs) to obtain the desired
properties. Therefore, the numerical evaluation relies on a truncation
procedure that is hard to justify \textit{a priori}.  The second program has
been highly successful in a number of cases
\cite{KLUMPER-SU3,JSUZUKI,JUTTNER,KLUMPER-HI,DAMERAU,RIBEIRO,OSP12}. Besides
the fact that no string hypothesis is required, the resulting nonlinear
integral equations are finite in number, which allows for an efficient
numerical evaluation. The drawback of  this approach is that no
systematic way to produce such sets of nonlinear integral equations has been
found so far, since one needs to define suitable auxiliary functions whose
construction can be very puzzling.

In this paper we apply the QTM method for $sl_n$-invariant models and propose
a set of auxiliary functions which are solutions to a finite set of NLIEs
(Section \ref{CANONICAL}). Our
choice is based on bilinear relations similar to the T-system
\cite{TSYSTEM,KUNIBA1998} with one additional advantage: we find factorization
of the symmetric fused terms, which implies the truncation  to a
  finite number of independent functions (Section \ref{GRAPHS}). We prove these relations through
manipulations of the Yangian version of Young tableaux (Section \ref{STAGE}).
In the special case of $sl_5$ (Section \ref{NUMERICS}), we present a numerical evaluation of
certain thermodynamical quantities for a model of hard core
  bosons resp.~itinerant pairs of electrons
on a chain.

\section{Setting the stage}\label{STAGE}

 \subsection{The Quantum Transfer Matrix}
The transfer matrix is a fundamental object in classical statistical
models. For a homogeneous system with degrees of freedom on the lattice one
can compute the partition function of the model by arranging its Boltzmann
weights into a matrix (Lax operator). Then, by performing simple algebraic
operations  like taking products of successive copies followed in
  case of periodic boundary conditions by a trace in the auxiliary space, one
  obtains the transfer matrix.  The largest eigenvalue of this yields the
  bulk properties of the model replacing the calculation of an infinite sum over all
  configurations. More specifically, in the case of two-dimensional lattice
models (such as vertex models, for example) one assigns independent degrees of
freedom to all bonds of the lattice, thus attaching a tensor product structure
to the Hilbert space. The Lax operator ${\cal L}$ is defined as
\eq
\mathcal{L}_{j,k}(\lambda,\nu)=\sum_{\alpha,\beta,\gamma,\delta} \check{\mathcal{L}}_{\alpha,\gamma}^{\beta ,\delta}(\lambda,\nu) e_{\alpha\beta}^{(j)} e_{\gamma\delta}^{(k)},
\en
where $e^{(j)}$ are tensorized Weyl matrices, acting only  on the
quantum (or vertical) space $V_j,\ j=1,\ldots,L$. Then, the transfer matrix is
given by
\eq
T(\lambda)=\tr_{\cal A} \prod_{i=1}^{\stackrel{\curvearrowleft}{L}} \mathcal{L}_{{\cal A}, i}(\lambda,0), \label{row-to-row}
\en
where we took the transfer direction from row to row and $V_{\cal A}$ denotes the auxiliary (or horizontal) space.

Integrability is realized by the commutativity of row-to-row
  transfer matrices. One way to see this is to construct a set of Boltzmann
weights parameterized by a (complex) spectral parameter such that
for arbitrary values $\lambda, \mu$
\eq
   [T(\lambda),T(\mu)]=0,
\label{comm}
\en
and hence $T(\lambda)$ serves as a generating family for the
  conserved charges through its derivative.  The latter condition is global,
while one only needs to know local Boltzmann weights to produce transfer
matrices. For Eq.\eqref{comm}
to hold a sufficient condition is the celebrated Yang-Baxter equation
\eq
{\cal L}_{12}(\lambda, \mu){\cal L}_{13}(\lambda, \gamma){\cal L}_{23}(\mu,
\gamma)={\cal L}_{23}(\mu, \gamma){\cal L}_{13}(\lambda, \gamma){\cal
  L}_{12}(\lambda, \mu). \label{YBaxter}
\en
Under the assumption that the model is fundamental, i.e.~all local
  spaces are isomorphic, we make no distinction between the intertwiner
($R$-matrix) and the Lax operator. Furthermore, for simplicity, in what
follows we will assume a set of desirable properties, namely
\begin{align}
  \mbox{Initial condition: } & {\cal L}_{12}(\lambda, \lambda)=P_{12}, \label{regul} \\
  \mbox{Time reversal: } & {\cal L}_{12}^{t_1}(\lambda, \mu)={\cal L}_{12}^{t_2}(\lambda, \mu), \label{time-rev}\\
  \mbox{Parity: } & \left[{\cal L}_{12}(\lambda, \mu),P_{12}\right]=0, \label{parity}
\end{align}
 where $P_{ij}$ denotes the permutation operator and $t_k$ the
  transposition in the $k$-th space.
A consequence of Eq.\eqref{YBaxter} in case of the standard initial condition
(regularity) for isomorphic spaces is the unitarity relation
\eq
   {\cal L}_{12}(\lambda,\mu) {\cal L}_{21}(\mu,\lambda) \propto \mbox{Id}. \label{uni}
\en
Also thanks to the standard initial condition, one can find a local Hamiltonian through the
logarithmic derivative of the row-to-row transfer matrix,
\eq
{\cal H}=\partial_{\lambda}\ln{T(\lambda)}\Big|_{\lambda=0} = \sum_{i=1}^L P_{i,i+1} \partial_{\lambda} {\cal L}_{i,i+1}\Big|_{\lambda=0}. \\
	\label{Hgen}
\en
A set of solutions satisfying all properties above corresponds to the Perk-Schultz model\cite{PERK}. The Lax operator in this case reads
\eq
{\cal L}(\lambda,\mu)= {\cal L}(\lambda -\mu,0)=(\lambda-\mu)\mbox{Id}+P,
\en
which satisfies
\eq
\left[{\cal L}_{12}(\lambda,\mu),{\cal G}_1+{\cal G}_2\right]=0,
\en
where ${\cal G}$ is the first fundamental representation of any generator of the $sl_n$ algebra.

In principle, one can use the Bethe ansatz to obtain the eigenvalues and the universal
eigenstates of the integrable manifold. However, in order to evaluate the
partition function of the spin chain one needs to sum over all
states. This task is best accomplished by following an alternative route.

The statistical operator $\ee^{-\beta {\cal H}}$ appears in the expansion of the row-to-row transfer matrix
\eq
T(\lambda)=\ee^{ \im {\cal P}+\lambda  \mathcal{H} +O(\lambda^2)}.
\label{texp}
\en
Now define a conjugated transfer matrix as
\eq
\bar{T}(\lambda)=\tr_{\cal A}{\prod_{i=1}^{\stackrel{\curvearrowleft}{L}} \mathcal{L}_{{\cal A},i}^{t_{\cal A}}(0,-\lambda) },
\label{transferBAR}
\en
so that it admits a similar expansion with the shift operator reversed
\eq
\bar{T}(\lambda)=\ee^{- \im {\cal P}+\lambda \mathcal{H} +O(\lambda^2)}.
\label{tbarexp}
\en
Consequently,
\begin{align}
	T(\lambda)\bar{T}(\lambda) = \ee^{2\lambda\mathcal{H} + O(\lambda^2)}.
\end{align}
By choosing $\lambda=-\frac{\beta J}{N}$, terms $O(\lambda^2)$ become small
compared to $O(\lambda)$ as we take the Trotter limit $N \rightarrow
\infty$. Therefore we find the Trotter-Suzuki decomposition,
\eq
Z=\lim_{N\rightarrow\infty}\tr\left[(T(-\tau)\bar{T}(-\tau))^{N/2} {\rm
    e}^{\beta \sum_{j=1}^{n}\mu_j \hat{N}_j}\right]=\tr ~{\rm e}^{-\beta
  (J \cal{H}-\mu \hat{N})},  \qquad \tau=\frac{\beta J}{N},
\label{particao}
\en
where we have included $n$ charges of the $sl_n$ algebra, here denoted by
$\hat{N}_j$, which  satisfy the constraint $\sum_{j=1}^n \hat{N}_j=L$. The
coupling constants $\mu_j$ play the role of generalized chemical potentials.

Note that the product $T(\lambda)\bar{T}(\lambda)$ may be understood as the transfer matrix of
a staggered lattice model. Instead of reading the products that make up the
partition function in the usual row-to-row direction one can read them from
column to column. This gives
rise to the quantum transfer matrix,
\eq
T^{QTM}(x)=\tr_{Q}\Bigg[{\rm e}^{\beta \sum_{k=1}^n \mu_k \hat{n}_{k,Q}} \prod_{i=1}^{\frac{N}{2}} ~ {\cal L}_{ 2i-1,Q}(-\tau, -\im x) {\cal L}_{ 2i,Q}^{t_Q}(-\im x,\tau)\Bigg].
\label{qtm-gen}
\en
The new spectral parameter $x$ is  introduced such that commutativity
holds $\left[T^{QTM}(x),T^{QTM}(y)\right]=0$. With this, the partition
function can be written as
\eq
Z=\lim_{N\rightarrow\infty}\tr ~ \left[{T^{QTM}(0)}\right]^{L}.
\en
Now the single largest eigenvalue,
$\Lambda_{\text{max}}^{QTM}(x)$, is needed to evaluate the thermodynamic
potential. Indeed,
\bear
f&=&-\frac{1}{\beta}\lim_{L,N\rightarrow \infty}\frac{1}{ L} \ln{Z}=-\frac{1}{\beta}\lim_{N\rightarrow \infty} \ln{\Lambda_{\text{max}}^{QTM}(0)}.
 \label{free-energy}
\ear

 \subsection{Yangian analogue of Young tableau}
The quantum transfer matrix inherits the integrability structure
  from its row-to-row counterpart. Both matrices possess the same intertwiner.  Nevertheless, because of the alternation that
defines Eq.(\ref{qtm-gen}), the reference states and the sectors associated to
$sl_n$ charges are different. Here
\eq
Q_k = \sum_{j=1}^{N} {(-1)}^j \hat{n}_{k,j},
\en
where $\hat{n}_{k,j}=0, 1$ depending on whether site $j$ is occupied by
species $k$ or not, i.e.~$\hat{n}_{k,j}$ are projectors onto
  orthogonal 1-dimensional spaces. On the account of $\sum_{k=1}^n
  \hat{n}_{k,j}=$id$_j$ we have the constraint $\sum_{k=1}^n Q_k = 0$ and
  hence not all $Q_k$ with $k=1,\ldots,n$ are independent.

 Consequently, in terms of Bethe ansatz expressions, only vacuum expectation
 values and sectors are changed. Choosing the reference state $\vert
 1,n,1,n,\ldots,1,n\rangle$, the eigenvalue is given by
{ \eq
\Lambda(x) = \sum_{j=1}^{n} \lambda_j(x),\label{Eigval}
\en}
 where the eigenvalue functions
 $\lambda_j(x),\ j=1,\ldots,n $ are given by \cite{{JUTTNER98}}
 \ytableausetup{centertableaux} \eq \lambda_j(x)= {\begin{ytableau} j
\end{ytableau}}_{~x}=
  \Phi_-(x) \Phi_+(x) \frac{q_{j-1}(x-\im) }{q_{j-1}(x)}\frac{q_j(x+\im)}{ q_j(x)} \ee^{\beta \mu_j}, \label{yangian}
\en
where
\eq
	q_j(x) = \begin{cases}
	\Phi_-(x), & j=0, \\
	\displaystyle\prod_{k=1}^{m_j} \left(x-x_{k}^{(j)}\right), & j=1,\ldots,n-1, \\
	\Phi_+(x), & j=n,
	\end{cases} \qquad  \Phi_{\pm}(x)=\left(x\pm \im \tau \right)^{N/2},
\en
and $x_{k}^{(j)},\ k=1,\ldots,m_j$ are solutions of the Bethe ansatz equations, given in the form
\begin{align}
	\lim_{x \rightarrow x_{k}^{(j)}} \frac{\lambda_j(x)}{\lambda_{j+1}(x)}=-1.\label{BAE}
\end{align}
We are specially interested in the sector
$m_1=m_2=\ldots=m_{n-1}=\frac{N}{2}$, which corresponds to $Q_k=0$ for all
$k$, where the largest eigenvalue is found.

In Eq.(\ref{yangian}) we made use of the Yangian version of the Young
tableau \cite{JSUZUKI94,KUNIBA95,KUNIBAOS95}. A rectangular
tableau of width $s$ and height $a$,
where $s,a$ are non-negative integers represents the product of
  functions (\ref{yangian})
\ytableausetup{mathmode,boxsize=2em,aligntableaux=center}
\eq
{\begin{ytableau}
j_{1,1} & j_{1,2} & j_{1,3} & \ldots & j_{1,s} \\
j_{2,1} & j_{2,2} & j_{2,3} & \ldots & j_{2,s} \\
\vdots & \vdots  &  \vdots  & \ddots & \vdots \\
j_{a,1} & j_{a,2} & j_{a,3} & \ldots & j_{a,s}
\end{ytableau}}_{~x}= \prod_{n=1}^a \prod_{m=1}^s
\ytableausetup{mathmode,boxsize=2em,aligntableaux=center}
{\begin{ytableau}
j_{n,m}
\end{ytableau}}_{~x +\im \left(n- \frac{a}{2}\right)-\im \left(m - \frac{s}{2}\right)}, \label{tableau}
\en
where the spectral parameter of the corresponding functions increases from top
to bottom and decreases from left to right on the tableau in
  steps of $\im$. It is important to observe the standard filling
rules when building a tableau, namely $j_{k,l} \leq j_{k,l+1}$ and $j_{k,l}<
j_{k+1,l}$.

Finally, note that the sum over all possible fillings of Eq.\eqref{tableau}
corresponds to the eigenvalue $\Lambda_{a,s}^{(n)}(x)$ of the fused transfer
matrix with anti-symmetric index $a=1,\ldots,n-1$ and symmetric index
$s=1,\ldots,\infty$.

\subsection{Deductive proof of T-system and fusion hierarchy}\label{PROOF}
Here we lay the groundwork to derive auxiliary functions in the forthcoming
sections. The auxiliary functions can be written conveniently in terms of
Young tableaux introduced above. They come in two sets, here denoted by
uppercase $\{B_j\}$ and lowercase $\{b_j\}$ letters and are related to each other by
 $B_j(x)=1+b_j(x)$. Our goal is to find nontrivial multiplicative functional relations between
the lowercase and uppercase functions. Our definition is inspired by the
well-known bilinear relations \cite{TSYSTEM} among fused eigenvalues, which go
by the name ``T-system''.

Although it is not our intention to discuss completeness of Bethe ansatz
eigenvectors, we may assume that at least a few eigenvalues are given in terms
of an analytical function in the manner we described above. In this sense, if
we find functional relations in terms of the Yangian analogue of the
Young tableaux, it is reasonable that some sectors of the transfer matrix if
not all do satisfy these functional relations, including the leading
eigenvalue. This is certainly the case if the relations are
  derived by, for example, the fusion algebra for operators.

We want to show the relation
\begin{align}
	T^{(a,s)}\left(x-\frac{\im}{2}\right) T^{(a,s)}\left(x+\frac{\im}{2}\right) = T^{(a-1,s)}\left(x\right) T^{(a+1,s)}\left(x\right)+T^{(a,s-1)}\left(x\right) T^{(a,s+1)}\left(x\right). \label{TSystemas}
\end{align}
As an example let us take two tableaux whose spectral parameters are shifted
from each other  by one in units of the crossing parameter. Let $a$ be the number of
rows and $s=1$ the number of columns of these tableaux. They are meant to
represent the product of $T^{(a, 1)}\left(x-\frac{\im}{2}\right) T^{(a,
  1)}\left(x+\frac{\im}{2}\right)$. From Eq.\eqref{TSystemas} it is clear that
in the case $a=4$ we would like to find
\begin{center}
\begin{align}	
\tikz {
	     \draw (1,.9) node {\begin{ytableau}
	    	\none[]\\
	    	*(green) \\
	    	\\
	    	\\
	    	\\
	    	\end{ytableau}};
    	\draw (1.8,-.9) node {\scriptsize $x+\frac{\im}{2}$};
         \draw (2.3,1) node {$\times$};
         \draw (3.4,.9) node {
			\begin{ytableau}
				\\
			*(green) \\
				\\
				\\
				\none[]
			\end{ytableau}};
	    \draw (4.2,-.2) node {\scriptsize $x-\frac{\im}{2}$};	
	    \draw (4.8,1) node {$=$};	
    	\draw (6,.9) node {\begin{ytableau}
    		\none[]\\
    		*(green) \\
    		\\
    		\\
    		\none[]
    		\end{ytableau}};
       \draw (6.5,-.25) node {\scriptsize $x$};
    	\draw (7,1) node {$\times$};
    \draw (7.9,.9) node {\begin{ytableau}
    		\\
    		*(green) \\
    		\\
    		\\
    		\\
    		\end{ytableau}};
    	\draw  (8.4,-.95) node {\scriptsize $x$};	
    	\draw (9,1) node {$+$};
       \draw (10.3,.55) node {\begin{ytableau}
       	*(green) ~ & ~ \\
       	~ & *(green) ~ \\
       	~ & ~\\
       	~ & ~
       	\end{ytableau}};
       \draw (11.15,-.95) node {\scriptsize $x$}
	}
 \label{ColumnTsystem}
\end{align}
\end{center}	
where  an empty tableau is implicitly understood as the sum over
all possible fillings. In the following we drop the explicit reference to the
spectral parameter whenever no ambiguity is possible. One way to do so is to
highlight at least the boxes that share one specific spectral parameter and
then
 preferably to align their horizontal positions, as exemplified
above. Before we jump
into Eq.\eqref{ColumnTsystem} \textit{per se} let us look into the most
fundamental fusion relation, which is actually the special case $a=s=1$. We
have
\begin{equation}
\begin{ytableau}
   \none[]~\\
   *(green)
\end{ytableau}
 \times
\begin{ytableau}
       \\
   \none[]~
\end{ytableau}=
\begin{ytableau}
    \\
*(green)
\end{ytableau}
+
\begin{ytableau}
  *(green) ~ &  ~
\end{ytableau}. \label{SimplestFusion}
  \end{equation}
Indeed, say the white and colored boxes on the left hand side of
Eq.(\ref{SimplestFusion}) are filled with any non-negative integer indices $j$
and $j_h$ respectively. Since either $j<j_h$ or $j_h \leq j$, which are
precisely the filling rules for Young tableaux of width 1 and height 1,
respectively, it is clear that the summation over all indices leads to the
right hand side of Eq.(\ref{SimplestFusion}).

Now let us consider the left hand side of Eq.(\ref{ColumnTsystem}). There are
several possibilities of rearranging the boxes according to their fillings,
for example
\begin{align}
\begin{ytableau}
\none[]\\
*(green)j_2\\
j_3\\
j_4\\
j_5\\
\end{ytableau} \times
\begin{ytableau}
i_1 \\
*(green)i_2 \\
i_3 \\
i_4 \\
\none[]
\end{ytableau}
=
\begin{ytableau}
\none[]\\
*(green)j_2\\
j_3\\
j_4\\
\none[]
\end{ytableau} \times
\begin{ytableau}
i_1\\
*(green)i_2\\
i_3\\
i_4\\
j_5\\
\end{ytableau}
&&\text{or}&&
\begin{ytableau}
\none[]\\
*(green)j_2\\
j_3\\
i_4\\
\none[]
\end{ytableau}\times
\begin{ytableau}
i_1\\
*(green) i_2\\
i_3\\
j_4\\
j_5\\
\end{ytableau}
&&\text{or}&&
\begin{ytableau}
\none[]\\
*(green)j_2\\
i_3\\
i_4\\
\none[]
\end{ytableau}\times \begin{ytableau}
i_1\\
*(green)i_2\\
j_3\\
j_4\\
j_5\\
\end{ytableau}
&&\text{or}&&
\begin{ytableau}
\none[]\\
*(green) i_2\\
i_3\\
i_4\\
\none[]
\end{ytableau}
\times
\begin{ytableau}
i_1\\
*(green)j_2\\
j_3\\
j_4\\
j_5\\
\end{ytableau}
&&\text{or}&&
\begin{ytableau}
*(green) j_2&i_1 \\
j_3& *(green)i_2 \\
j_4&i_3 \\
j_5&i_4
\end{ytableau}, \label{FusionMoves}
\end{align}
which correspond, respectively, to the conditions:\newline
$~{\rm i})~j_5>i_4$\newline
$~{\rm ii})~j_5\leq i_4,~~~j_4>i_3$ \newline
$~{\rm iii})~j_5\leq i_4,~~~j_4\leq i_3,~~~j_3>i_2$ \newline
$~{\rm iv})~j_5\leq i_4,~~~j_4\leq i_3,~~~j_3\leq i_2,~~~j_2>i_1,$\newline
$~{\rm v})~j_5\leq i_4,~~~j_4\leq i_3,~~~j_3\leq i_2,~~~j_2 \leq i_1$.\newline

Note conditions i)--iv) satisfy all allowed filling rules for a combination of
two column tableaux of sizes 3 and 5, i.e.~the first term on the RHS of
Eq.\eqref{ColumnTsystem}, while condition v) describes exactly the filling of
a tableau of width 2 and height 4, i.e.~the second term on the RHS of
Eq.\eqref{ColumnTsystem}. What is left to show is that conversely any term
appearing in the first term on the RHS of Eq.\eqref{ColumnTsystem} appears in
one of the cases i)--iv). This can be done which we leave to the
reader. This shows that the sum of the cases i)--iv) amounts to
  the first term on the RHS of Eq.\eqref{ColumnTsystem} where each of the two
  columns can be summed over independently.  This completes the proof of
identity Eq.\eqref{FusionMoves}. One can apply the same reasoning recursively to
show Eq.\eqref{TSystemas} for arbitrary values of $a,\ s$.

In the Yangian analogue of the Young tableau, boxes can only exchange positions when they have the same spectral parameter. This helps to pinpoint the possible outcomes in relations like the above. For instance, on the basis of the same analysis we may find
\begin{equation}
\begin{ytableau}
    *(green) & ~&~ \\
    ~ & \none[]&\none[]
    \end{ytableau} \times
    \begin{ytableau}
    *(green)
    \end{ytableau}=
    \begin{ytableau}
    *(green)~ & ~&~ \\
    ~ & *(green)&\none[]
    \end{ytableau}+
    \begin{ytableau}
    *(green)\\
    ~\end{ytableau} \times
    \begin{ytableau}
    *(green)~ & ~&~     \end{ytableau}.
\end{equation}
It is then a pleasant game to use these ``fusion moves'' in order to derive functional relations among transfer matrices.

\section{Adjacency matrices}\label{GRAPHS}

There are circumstances where the set of functional relations provided by the T-system truncates by itself. This can be viewed as a generalization of the
inversion identity \cite{BAZHANOV,TSYSTEM}, which can either be used to derive
finite sets of NLIE or to describe the spectrum in terms of the zeros of
the eigenvalue functions. Nevertheless, in the general case such a ``spontaneous'' truncation does not
take place, which makes it more natural to represent spectral
  values in terms of (fused) eigenvalues and Bethe ansatz, where extra
polynomials like the $q_j(x)$ functions also enter explicitly. More generally, other types
of functions may appear in different sets of functional equations
  for suitable auxiliary functions based on Young diagrams with
    restricted filling rules.
In this section, we obtain these
functions, henceforth called {\unknown}s
 for elementary analytic functions or elementary analytic factors.

\subsection{Analytical consequences of Bethe ansatz for fundamental representations}
 The eigenvalue Eq.(\ref{Eigval}) as function of the spectral parameter may have
 poles at points related to the zeros of the functions $q_j(x)$. Since the
 eigenvectors of the transfer matrix do not depend on the spectral parameter,
 the corresponding eigenvalues like all matrix elements must be analytical
 functions of $x$. This way one recovers the Bethe ansatz
 equations Eq.(\ref{BAE}) whose zeros are those of the
 $q_j(x),\ j=1,\ldots,n-1$. For the first fundamental representation ($a=1$)
 the removability of these poles happens pairwise through the sum
 $\lambda_j(x)+\lambda_{j+1}(x)$. On the other hand, for the cases in which
 the sum of tableaux is truncated,  due to restricted filling
   rules, not all poles are removed. To see this, one must first study
 case-by-case the relationship between any two tableaux that belong to the
 same Young diagram resp. representation and check whether they
 are connected via a Bethe ansatz equation. Then we encode this information
 into so-called adjacency matrices from which later we define the additional
 {\unknown}s that appear in the definition of the auxiliary functions with the
 proper poles.
	
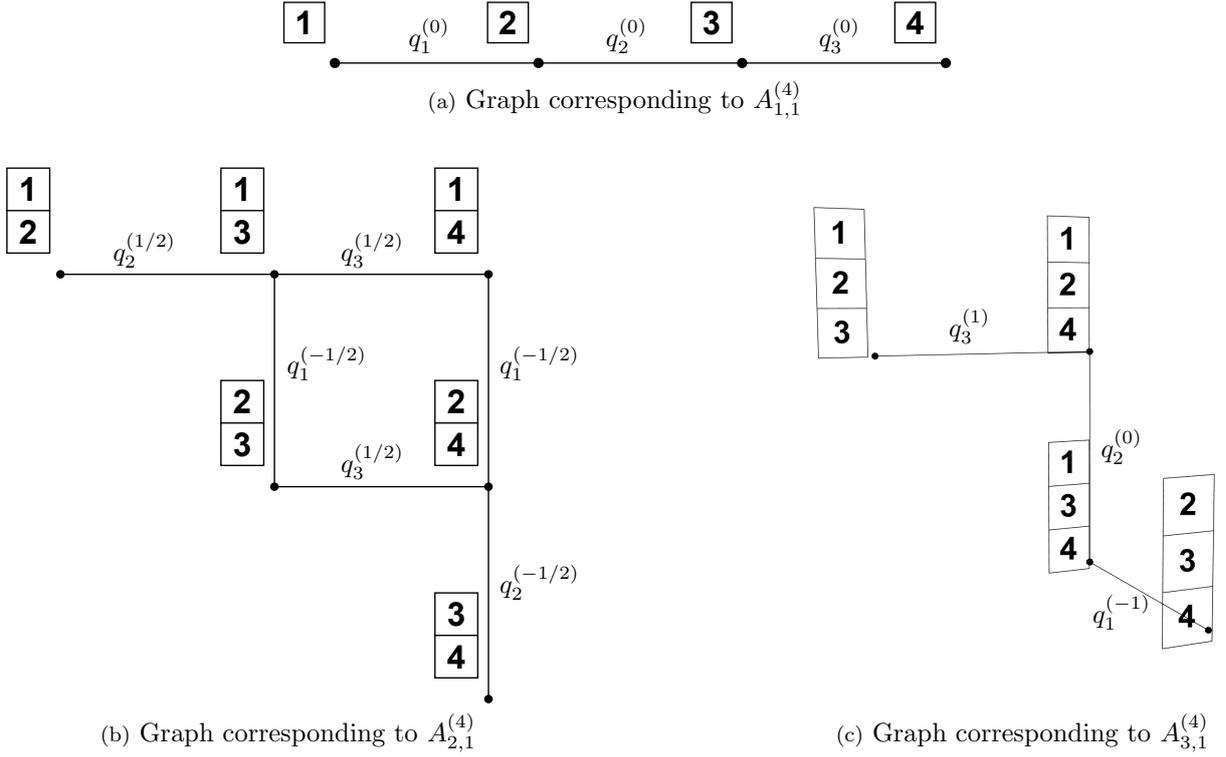
\begin{figure}[h!]
	\centering
	\subfigure[\normalsize Graph corresponding to $A_{1,1}^{(4)}$]{\begin{tikzpicture}
		\node[anchor=south west,inner sep=0] at (0,0) {\includegraphics[width=.55\textwidth]{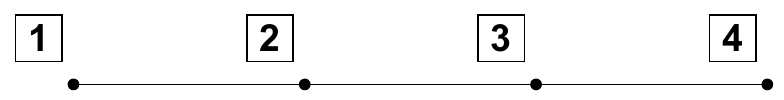}};
		\node at (2.1,0.55) {$q_1^{(0)}$};
		\node at (4.7,0.55) {$q_2^{(0)}$};
		\node at (7.5,0.55) {$q_3^{(0)}$};
		\end{tikzpicture}\label{sl4-1st}}
	\subfigure[\normalsize Graph corresponding to $A_{2,1}^{(4)}$]{
		\begin{tikzpicture}
		\node[anchor=south west,inner sep=0] at (0,0) {\includegraphics[width=.4\textwidth]{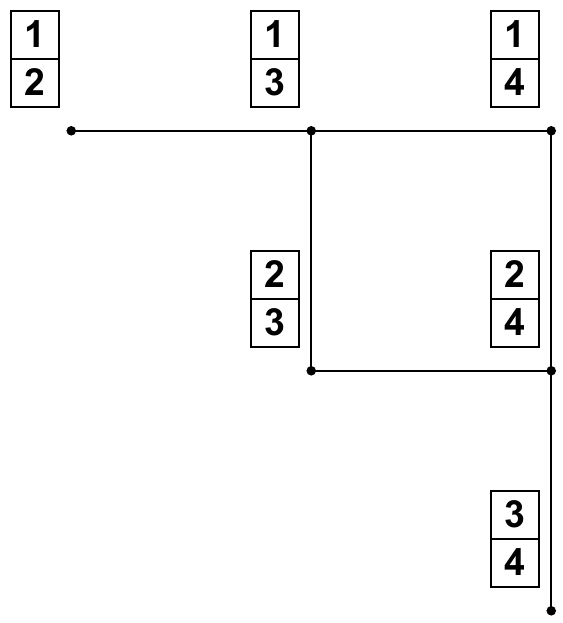}};
		\node at (1.95,6.1) {$q_2^{(1/2)}$};
		\node at (4.95,6.1) {$q_3^{(1/2)}$};
		\node at (4.95,3.3) {$q_3^{(1/2)}$};
		\node at (4.35,4.6) {$q_1^{(-1/2)}$};
		\node at (7.15,4.6) {$q_1^{(-1/2)}$};
		\node at (7.15,1.7) {$q_2^{(-1/2)}$};
		\end{tikzpicture}\label{sl4-2nd}}
	\hfill
	\subfigure[\normalsize Graph corresponding to $A_{3,1}^{(4)}$]{\begin{tikzpicture}
		\node[anchor=south west,inner sep=0] at (0,0) {\includegraphics[width=.35\textwidth]{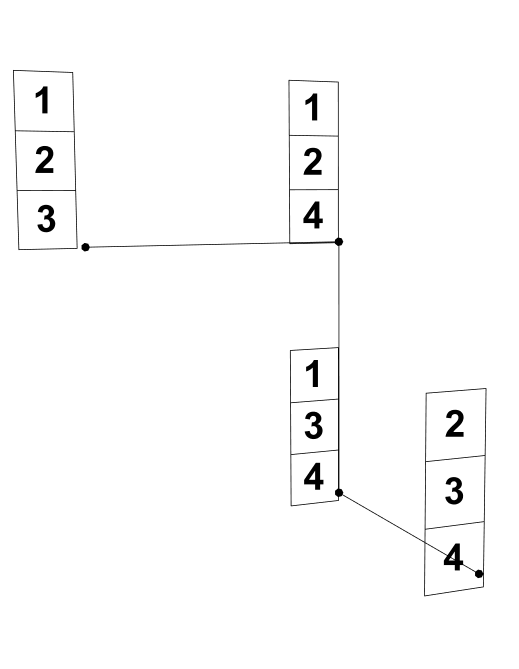}};
		\node at (2.2,5.1) {$q_3^{(1)}$};
		\node at (4.2,3.5) {$q_2^{(0)}$};
		\node at (4.2,1.3) {$q_1^{(-1)}$};
		\end{tikzpicture}\label{sl4-3rd}}
		\caption{Pole removal graphs for the fundamental representations of $sl_4$. The $q$-functions on the bonds refer to the shared pole between the vertices it connects. We use the notation: $q_j^{(k)} \equiv q_j(x+\im k) $.}
		\label{sl4-graphs}
\end{figure}

The adjacency matrices are represented by graphs where each vertex corresponds
to a term of the eigenvalue $\Lambda_{a,1}^{(n)}$ and each edge to the shared
pole between the vertices.  In Fig.\ref{sl4-graphs} we present the graphs of
$sl_4$ with $s=1$ and $a=1, 2$ as examples. For the first fundamental representation the graph would
simply amount to the ${\cal A}_{n-1}$ Dynkin diagram. See Fig.\ref{sl4-1st}. The
situation becomes more intricate as we move to fundamental representations
with $a>1$.

\ytableausetup{mathmode,boxsize=1em,aligntableaux=center}

While a graphical approach consisting in identifying the proper {\unknown}s
from cuts of these graphs is possible, for representations with
  $a>3$ this would become unfeasible on the account that these graphs spread
  to higher dimensions. To this end the adjacency matrices are more
versatile. Nevertheless we refer to Appendix \ref{app-graphs} in which we show
how the graphs are used for the $sl_4$ case.

 We enumerate the vertices by ordering them resp.~the
  corresponding tableaux such that the tableau with smaller indices precedes the rest. For
instance, in the case of the $a=2$ representation of $sl_4$ we set
	\ytableausetup{boxsize=1.5em}
	\begin{align}
		\bt 1 \\ 2 \et \prec 	\bt 1 \\ 3 \et \prec 	\bt 1 \\ 4 \et \prec 	\bt 2 \\ 3 \et \prec 	\bt 2 \\ 4 \et \prec 	\bt 3 \\ 4 \et.
	\end{align} \label{hierarchy}
Having established this ordering, set the entries of the
  adjacency matrix at position $(i,j)$ to $0$ if the
vertices No.~$i$ and No.~$j$ are not connected or to the corresponding $q_k(x)$ of the removed set of
poles if they are connected. These matrices have to be
  worked out individually. As an example, the matrices for all fundamental
  representations of the $sl_4$ invariant model with $s=1$ are given by
\begin{subequations}
	\begin{align}
		A_{1,1}^{(4)}(x) &= \begin{bmatrix}
		0 & q_1^{(0)} & 0 & 0 \\
		q_1^{(0)} & 0 & q_2^{(0)} & 0 \\
		0 & q_2^{(0)} & 0 & q_3^{(0)} \\
		0 & 0 & q_3^{(0)} & 0
		\end{bmatrix}, \label{adj-sl4-1} \\
		A_{2,1}^{(4)}(x) &= \begin{bmatrix}
		0 & q_2^{(1/2)} & 0 & 0 & 0 & 0 \\
		q_2^{(1/2)} & 0 & q_3^{(1/2)} & q_1^{(-1/2)} & 0 & 0 \\
		0 & q_3^{(1/2)} & 0 & 0 & q_1^{(-1/2)} & 0 \\
		0 & q_1^{(-1/2)} & 0 & 0 & q_3^{(1/2)} & 0 \\
		0 & 0 & q_1^{(-1/2)} & q_3^{(1/2)} & 0 & q_2^{(-1/2)} \\
		0 & 0 & 0 & 0 & q_2^{(-1/2)} & 0
		\end{bmatrix}, \label{adj-sl4-2} \\
		A_{3,1}^{(4)}(x) &= 	\begin{bmatrix}
		0 & q_3^{(1)} & 0 & 0 \\
		q_3^{(1)} & 0 & q_2^{(0)} & 0 \\
		0 & q_2^{(0)} & 0 & q_1^{(-1)} \\
		0 & 0 & q_1^{(-1)} & 0
		\end{bmatrix}. \label{adj-sl4-3}
	\end{align}
\end{subequations}

\subsection{Identification of {\unknown}s and notation}
As we can see from the graphs, the Bethe equations imply
  analyticity for any eigenvalue function, i.e.~for the sum over all terms
  represented by a certain Young tableau. But also partial sums over such
  terms show the phenomenon of pole cancellation due to the Bethe equations,
  however with certain poles not being  removed. The partial sums appear in
  factorizations of the later to be introduced auxiliary functions. It will be
  important to write the partial sums in terms of analytical functions divided
  by products of $q$-functions with certain shifts of the arguments. The use
  of the Fourier transform for the logarithmic derivatives of all functional
  equations will then allow us to derive non-linear integral equations for the
  auxiliary functions.  In general, one may define the analytical factor
  functions by summing over tableaux, checking from the adjacency matrix which
  poles are being removed and which are not and establishing the combination
  \ytableausetup{boxsize=1.5em}
\begin{equation}
\frac{\prod_{\alpha \in \text{non removed poles}} q_{\ind(\alpha)}(x+\im \sft(\alpha) )}{\prod_{\beta \in \text{common zeros}} \Phi_{\sg(\beta)} (x+\im \sft(\beta))}   \sum_{\{j\}} \begin{ytableau} j_1 \\ \vdots \\ j_n \end{ytableau}   = p(x), \label{part}
\end{equation}
as a polynomial function $p(x)$.  Here, certain index, shift and sign
functions $\ind : \alpha \rightarrow 1,~2,\ldots,~n$, $\sft : \alpha
\rightarrow \mathbb{Z}/2$ and $\sg : \beta \rightarrow \pm$ appear. The label
$\alpha$ indexes the set of poles of individual terms without cancellation in
the sum, $\beta$ indexes the factors of type $\Phi_{\pm}$ appearing jointly in
all terms of the sum thanks to definition Eq.\eqref{yangian}. The functions
$\ind$ and $\sg$ denote the type of poles and zeros, the function $\sft$ gives
the shift of the arguments of the elementary factors. Then, the nonremoved
poles -- namely the ones appearing as elements of the  adjacency
  matrices in rows of any term of the sum in question, but in a column of a
term not appearing in that sum -- result in the numerator of Eq.\eqref{part}.
The factors $\Phi_-$ and $\Phi_+$ appearing in the denominator originate from
factoring out $q_0 = \Phi_-$ or $q_n = \Phi_+$ jointly appearing in
all terms of the sum. The eigenvalues $\Lambda_{a,1}^{(n)}(x)$ are also
{\unknown}s, obtained when we consider the full matrix. Of course, in this case
there are no poles left behind.

For instance, take the first fundamental representation of
$sl_4$. The sum $\begin{ytableau} 1 \end{ytableau}+\begin{ytableau}
2 \end{ytableau}$ removes the poles at $q_1(x)$ but keeps the poles at
$q_2(x)$ (see $A_{1,1}^{(4)}(x)$) above. Moreover common trivial zeros may be factorized
and one can show that
\begin{equation}
\frac{q_2(x)}{\Phi_+(x)} \left(\begin{ytableau} 1 \end{ytableau}+\begin{ytableau} 2 \end{ytableau}\right) =: p_{1,1}^{(4)}(x),
\end{equation}
is an entire function. More specifically, it is a polynomial of degree
$\frac{N}{2}+m_2$. On the other hand, as an example for the second representation, we may use $A_{2,1}^{(4)}(x)$ to obtain
\begin{align}
	\frac{q_2\left(x-\frac{\im}{2}\right)}{\Phi_+\left(x-\frac{\im}{2}\right)\Phi_-\left(x+\frac{\im}{2}\right)}\left(\bt 1 \\ 2 \et + \bt 1 \\ 3 \et + \bt 1 \\ 4 \et + \bt 2 \\ 3 \et + \bt 2 \\ 4 \et\right) =:  p_{2,2}^{(4)}(x).
\end{align}
It is worth noting that representation matrices  with larger $a$ may give the  already obtained {\unknown}s as well as new ones. Also, the matrix $A_{n-1,1}^{(n)}(x)$ does not contribute with new {\unknown}s for the $sl_n$ case except for $\Lambda_{n-1,1}^{(n)}(x)$.

The polynomials $p_{a,k}^{(n)}(x)$ together with $q_{1}(x),~\ldots,q_n(x)$ and
the eigenvalues $\Lambda_{a,1}^{(n)}$  will be called the \textit{{\unknown}s of the system}.
They are blocks in terms of which we will describe the (bi)linear
relations underlying the definition of auxiliary functions. In other words, lowercase and uppercase
auxiliary functions (see the definition in the first paragraph of
Sect.~\ref{PROOF}) are to be factorized solely in terms of these blocks. They
are completely defined through the partial sum Eq.(\ref{part}) over Young
tableaux.

Before we continue, we set a convenient
notation for the {\unknown}s. Let $\{i_1,~i_2,\ldots,~i_a\}$,
$\{j_1,~j_2,\ldots,~j_a\}$ be the sets of indices corresponding to the initial
and final tableaux, respectively, in the ordering explained in Eq.\eqref{hierarchy}.
$i_{\ell} \leq j_{\ell}$ for $\ell=1,\ldots,a$. Then we set
\ytableausetup{boxsize=2.5em}
{\begin{equation}
	\begin{ytableau}
	i_1,j_1\\
	i_2,j_2\\
	\vdots\\
	i_a,j_a
	\end{ytableau} =
	\ytableausetup{boxsize=2.5em}
	\sum_{\{i_{\ell}\leq k_{\ell}\leq j_{\ell}\}}\begin{ytableau}
	k_1\\
	k_2\\
	\vdots\\
	k_a
	\end{ytableau}
	=\begin{ytableau}
	i_1\\
	i_2\\
	\vdots\\
	i_a
	\end{ytableau}
	+ \ldots+
	\begin{ytableau}
	j_1\\
	j_2\\
	\vdots\\
	j_a
	\end{ytableau}, \label{notation}
	\end{equation}}
where the dependence on the spectral parameter has been omitted and the sum over $k_{\ell}$ satisfies the standard filling rule. This is a good
{\unknown} for a Young tableau corresponding to the fundamental representation $a$
if no factorization takes place, which happens for $j_{\ell} \geq i_{\ell+1}$ with $\ell = 1,~2,\ldots,~n-1$. These are the column {\unknown}s. It is not hard to see how this generalizes to
other kinds of tableaux.

\section{A canonical set of functions}\label{CANONICAL}

\subsection{Prescription from fusion}
For the $sl_n$ invariant model in the $a$-th fundamental representation the T-system reads
\begin{equation}
T^{(a,1)}\left(x-\frac{\im}{2}\right) T^{(a,1)}\left(x+\frac{\im}{2}\right) = T^{(a-1,1)}\left(x\right) T^{(a+1,1)}\left(x\right)+T^{(a,0)}\left(x\right) T^{(a,2)}\left(x\right). \label{TSystemk}
\end{equation}
From Eq.\eqref{TSystemk} the Y-system is defined as, see e.g.~\cite{JUTTNER98}
\begin{align}
	Y^{(a,1)}(x)=\frac{T^{(a,1)}\left(x-\frac{\im}{2}\right) T^{(a,1)}\left(x+\frac{\im}{2}\right)}{T^{(a-1,1)}\left(x\right) T^{(a+1,1)}\left(x\right)}, && 	y^{(a,1)}(x)=\frac{T^{(a,0)}\left(x\right) T^{(a,2)}\left(x\right)}{T^{(a-1,1)}\left(x\right) T^{(a+1,1)}\left(x\right)}, \label{YSystem}
\end{align}
thus $Y^{(a,1)}(x) = 1+y^{(a,1)}(x)$. As mentioned before we want to generate
sets of auxiliary functions that are  related to this very same expression. One
way to achieve this is to modify the filling rules for the tableaux of
Eq.\eqref{YSystem} so that not all possibilities are allowed at once -- as a
consequence, instead of having solely the eigenvalues $\Lambda^{(a,s)}(x)$ in
the definition of these auxiliary functions they will be constructed mainly
with the {\unknown}s introduced in the last section, which opens up a number of
new combinations.

Let $j_1<\ldots<~j_a$, $j_k=1,\ldots,n$ be fixed numbers that act as boundaries
for the indices in the boxes on the left hand side of Eq.(\ref{TSystemk}). In
order to fill up a column tableau from top to bottom we allow only a certain
range of fillings determined according to its spectral parameter. More
specifically,
\ytableausetup{boxsize=2.5em}
\begin{align}
	x-\frac{a\im}{2}+\frac{\im}{2} \quad  \rightarrow \quad	\begin{ytableau}
	1,j_1
	\end{ytableau},  &&	x-\frac{a\im}{2}+\frac{3\im}{2} \quad \rightarrow \quad	\begin{ytableau}
	j_1,j_2
	\end{ytableau}, && \ldots && x+\frac{a\im}{2}+
        \frac{\im}{2} \quad \rightarrow \quad	\begin{ytableau}
	j_a,n
	\end{ytableau}.
\end{align}
where we used the compact notation introduced in Eq.\eqref{notation}. As an example, take $n\geq 5$ and $a=4$ and apply these filling rules to the tableaux on the left hand side of Eq.\eqref{TSystemk}. By means of the ``fusion moves'' explained in Sec.\ref{PROOF} the rearranging of boxes yields  \ytableausetup{boxsize=2.5em}

\begin{align}
\begin{ytableau}
   \none[]
    \\
    j_1,j_2\\
    j_2,j_3\\
    j_3,j_4\\
    j_4,n
\end{ytableau} && \times && \begin{ytableau}
    1,j_1\\
    j_1,j_2\\
    j_2,j_3\\
    j_3,j_4
    \\
   \none[]
\end{ytableau}
 && = &&
 \begin{ytableau}
   \none[]\\
    j_1,j_2\\
    j_2,j_3\\
    j_3,j_4\\
   \none[]
\end{ytableau}
 && \times &&
\begin{ytableau}
    1,j_1\\
    j_1,j_2\\
    j_2,j_3\\
    j_3,j_4\\
    j_4,n
\end{ytableau}
&& + &&
\begin{ytableau}
    j_1,j_2&1,j_1 \\
    j_2,j_3&j_1,j_2 \\
    j_3,j_4&j_2,j_3 \\
    j_4,n&j_3,j_4
\end{ytableau} && = &&
\begin{ytableau}
   \none[]\\
    j_1,j_2\\
    j_2,j_3\\
    j_3,j_4\\
   \none[]
\end{ytableau}
&& \times &&
\begin{ytableau}
    1,j_1\\
    j_1,j_2\\
    j_2,j_3\\
    j_3,j_4\\
    j_4,n
\end{ytableau}
&& + &&
\begin{ytableau}
    j_1&j_1 \\
    j_2&j_2 \\
    j_3&j_3 \\
    j_4&j_4
\end{ytableau},
\label{DefinitionCanonicalk4}
\end{align}
Observe that the tableaux in Eq.(\ref{DefinitionCanonicalk4}) in general are
not further reduced in the sense that there might be fillings of the
individual boxes precluded by the filling rules of the Young tableau. For
instance, if one takes $j_1=1$ then of course one cannot start the following
box with the value $1$. Instead, one should start with value $2$, even though we indicate
there that it may start from $j_1=1$. Therefore, reduction is
circumstantial to the choice of the $j_k$. On the other hand, the symmetrically
fused tableau on the right hand side reduces completely regardless the choice
of $j_k$. Our construction transforms the sum in the symmetrically-fused
tableau into a single term which can be factorized in terms of column
tableaux. This truncation is important to avoid indefinite growth of {\unknown}s
(and therefore of auxiliary functions) thus making the NLIEs finite in
number. In fact, for a $sl_n$-symmetric model our approach provides $2^n-2 =
\sum_{a=1}^{n-1} \binom{n}{a}$ uppercase and lowercase auxiliary
functions. This is also the number of {\unknown}s that naturally appear in this
construction. Hence, as long as proper domains of the complex plane can be chosen for all {\unknown}s,
they can be eliminated in Fourier space thus providing a consistent set of
NLIEs.

In the general case the auxiliary functions are given as
\ytableausetup{boxsize=3.2em}
\begin{align}
{B}_{a,j}^{(n)}(x)=
\frac{
	\begin{ytableau}
	1,j_1\\
	j_1,j_2\\
	\vdots\\
	j_{a-1},j_a
	\\
	\none[]
	\end{ytableau} \times
	\begin{ytableau}
	\none[]
	\\
	j_1,j_2\\
	\vdots\\
	j_{a-1},j_a\\
	j_a,n
	\end{ytableau}
}
{
	\begin{ytableau}
	1,j_1\\
	j_1,j_2\\
	\vdots\\
	j_{a-1},j_a\\
	j_a,n
	\end{ytableau} \times
	\begin{ytableau}
	\none[]\\
	j_1,j_2\\
	\vdots\\
	j_{a-1},j_a\\
	\none[]
	\end{ytableau}
}, && \qquad {b}_{a,j}^{(n)}(x)=
\frac{
	\begin{ytableau}
	j_1&j_1 \\
	j_2&j_2 \\
	\vdots&\vdots \\
	j_a&j_a
	\end{ytableau}
}
{
	\begin{ytableau}
	1,j_1\\
	j_1,j_2\\
	\vdots\\
	j_{a-1},j_a\\
	j_a,n
	\end{ytableau} \times
	\begin{ytableau}
	\none[]\\
	j_1,j_2\\
	\vdots\\
	j_{a-1},j_a\\
	\none[]
	\end{ytableau}
},
\end{align}
\normalsize
with ${B}_{a,j}^{(n)}(x) = 1+b_{a,j}^{(n)}(x)$, where $j$ here is a vector index that follows the same hierarchy as exemplified in Eq.\eqref{hierarchy}, i.e.~ $a=1\Rightarrow j \equiv j_1 = 1,2,\ldots,n$,  $a=2 \Rightarrow j \equiv (j_1,j_2) = (1,2), (1,3),\ldots,(3,4)$, and so on.

\subsection{NLIE for $sl_4$ and $sl_5$}\label{can-sl45}
In this section we will apply the canonical construction in order to obtain
auxiliary functions for $sl_4$- and $sl_5$-symmetric models. We separate the
functions into sets corresponding to each fundamental representation of
$sl_n$, which is labeled by the index $a=1,\ldots,n-1$. Each representation
contains $d_a^{(n)}=\binom{n}{a}$ uppercase and lowercase auxiliary functions, which corresponds to its dimension. One can
verify that all functions in a given representation $a=k$ are conjugated to
the ones in $a=n-k$. Also, each representation is invariant by species
conjugation, i.e.~the change in indices $k \leftrightarrow n-k+1$, etc. It is
worth noting that the auxiliary functions of $sl_2$ and $sl_3$ presented in
Appendix \ref{prev} can be obtained through this approach. Later on, we
derive sets of NLIEs involving these auxiliary functions.

\subsubsection{$sl_4$}
Using the canonical prescription, one finds 14 auxiliary functions for the $sl_4$
case. We observe that 10 of them coincide with the result of Damerau
and Kl\"{u}mper  \cite{DAMERAU}. In the first representation, $a=1$, one finds
\ytableausetup{boxsize=2em}
\begin{subequations}
	\begin{align}
	B_{1,1}^{(4)}(x)&=\frac{\bt 1,4\et_{x+\im/2}}{\bt 2,4 \et_{x+\im/2}}, & b_{1,1}^{(4)}(x)&=\frac{\bt 1 \et_{x+\im/2}}{\bt 2,4 \et_{x+\im/2}},\label{b4-11} \\
	B_{1,2}^{(4)}(x)&=\frac{\bt 1 , 2 \et_{x-\im/2} \bt 2, 4 \et_{x+\im/2}}{\bt 1,2 \\ 2,4 \et_x}, & b_{1,2}^{(4)}(x)&=\frac{\bt 2 \et_{x-\im/2} \bt 2 \et_{x+\im/2}}{\bt 1,2 \\ 2,4 \et_x},\label{b4-12}\\
	B_{1,3}^{(4)}(x)&=\frac{\bt 1,3 \et_{x-\im/2}\bt 3,4 \et_{x+\im/2}}{\bt 1,3 \\ 3,4\et_x}, & b_{1,3}^{(4)}(x)&=\frac{\bt 3 \et_{x-\im/2} \bt 3 \et_{x+\im/2}}{\bt 1,3 \\ 3,4 \et_x},\label{b4-13} \\
	B_{1,4}^{(4)}(x)&=\frac{\bt 1,4 \et_{x-\im/2}}{\bt 1,3 \et_{x-\im/2}}, & b_{1,4}^{(4)}(x)&=\frac{\bt 4 \et_{x-\im/2}}{\bt 1, 3 \et_{x-\im/2}},\label{b4-14}
	\end{align}
\end{subequations}
followed by the self-conjugated set of functions for $a=2$,
\begin{subequations}
	\begin{align}
	B_{2,1}^{(4)}(x)&=\frac{\bt 1,2 \\ 2,4 \et_{x+\im/2}}{\bt 1,2 \et_x \bt 3,4 \et_{x+\im}}, & b_{2,1}^{(4)}(x)&=\frac{\bt 1 \\ 2 \et_{x+\im/2}}{\bt 1,2 \et_x \bt 3,4 \et_{x+\im}},\label{b4-21}\\
	B_{2,2}^{(4)}(x)&=\frac{\bt 2,3 \et_x \bt 1,3 \\ 3,4 \et_{x+\im/2}}{\bt 1,3\et_x \bt 2,3 \\ 3,4 \et_{x+\im/2}} , & b_{2,2}^{(4)}(x)&=\frac{\bt 3 \et_x \bt 1 \\ 3 \et_{x+\im/2}}{\bt 1,3 \et_x \bt 2,3\\3,4 \et_{x+\im/2}} ,\label{b4-22} \\
	B_{2,3}^{(4)}(x)&=\frac{\bt 1,3 \et_x \bt 2,4 \et_x }{\bt 1,4 \et_x \bt 2,3 \et_x}, & b_{2,3}^{(4)}(x)&=\frac{\bt 1 \et_x  \bt 4 \et_x }{\bt 1,4 \et_x \bt 2,3 \et_x },\label{b4-23} \\
	B_{2,4}^{(4)}(x)&=\frac{\bt  2,3 \\ 3,4 \et _{x+\im/2} \bt 1,2 \\ 2,3  \et_{x-\im/2}}{\bt  2,3  \et_x \bt 1,2 \\ 2,3 \\ 3,4 \et_x}, & b_{2,4}^{(4)}(x)&=\frac{\bt 2 \\ 3 \et_{x+\im/2}  \bt 2 \\ 3  \et_{x-\im/2}}{\bt  2,3 \et_x \bt 1,2 \\ 2,3 \\ 3,4 \et_{x}},\label{b4-24} \\
	B_{2,5}^{(4)}(x)&=\frac{\bt 2,3 \et_x \bt 1,2 \\ 2,4 \et_{x-\im/2}}{\bt 2,4 \et_x \bt 1,2 \\ 2,3 \et_{x-\im/2}} , & b_{2,5}^{(4)}(x)&=\frac{\bt 2 \et_x \bt 2 \\ 4 \et_{x-\im/2}}{\bt 2,4 \et_x\bt 1,2 \\ 2,3 \et_{x-\im/2}},\label{b4-25} \\
	B_{2,6}^{(4)}(x)&=\frac{\bt 1,3 \\ 3,4 \et_{x-\im/2}}{\bt 1,2 \et_{x-\im} \bt 3,4 \et_{x-\im/2}}, &
	b_{2,6}^{(4)}(x)&=\frac{\bt 3 \\ 4 \et_{x-\im/2}}{\bt 1,2 \et_{x-\im} \bt 3,4 \et_{x}},\label{b4-26}
	\end{align}
\end{subequations}
and, finally, conjugated to the first representation, the functions corresponding to $a=3$ are
\begin{subequations}
	\begin{align}	
	B_{3,1}^{(4)}(x)&=\frac{\bt 1,2 \\ 2,3 \\ 3,4 \et_{x+\im/2}}{\bt 1,2 \\ 2,3 \\ 4 \et_{x+\im/2}}, & b_{3,1}^{(4)}(x)&=\frac{\bt 1 \\ 2 \\ 3 \et_{x+\im/2}}{\bt 1,2 \\ 2,3 \\ 4 \et_{x+\im/2}},\label{b4-31} \\
	B_{3,2}^{(4)}(x)&=\frac{\bt 3,4 \et_{x+\im/2} \bt 1,2 \\ 2,3 \et_x}{\bt 3 \et_{x+\im/2} \bt 1,2 \\ 2,4 \et_x}, & b_{3,2}^{(4)}(x)&=\frac{\bt 4 \et_{x+\im/2}\bt 1 \\ 2 \et_x}{\bt 3 \et_{x+\im/2}\bt 1,2 \\ 2,4 \et_x} ,\label{b4-32} \\
	B_{3,3}^{(4)}(x)&=\frac{\bt 1,2 \et_{x-\im/2} \bt 2,3 \\ 3,4 \et_x}{\bt 2 \et_{x-\im/2} \bt 1,3 \\ 3,4 \et_x} , & b_{3,3}^{(4)}(x)&=\frac{\bt 1 \et_{x-\im/2}}{\bt 2 \et_{x-\im/2}} \frac{\bt 3 \\ 4 \et_x}{\bt 1,3 \\ 3,4 \et_x},\label{b4-33} \\
	B_{3,4}^{(4)}(x)&=\frac{\bt 1,2 \\ 2,3 \\ 3,4 \et_{x-\im/2}}{\bt 1 \\ 2,3 \\ 3,4 \et_{x-\im/2}}, &
	b_{3,4}^{(4)}(x)&=\frac{\bt 2 \\ 3 \\ 4 \et_{x-\im/2}}{\bt 1 \\ 2,3 \\ 3,4 \et_{x-\im/2}}.\label{b4-34}
	\end{align}
\end{subequations}

Now we want to replace the tableaux by the functions they represent. In the $sl_4$ case we have three eigenvalue expressions, namely
\ytableausetup{boxsize=2em}
\begin{align}
\Lambda_{1,1}^{(4)}(x)&=\bt 1,4 \et_x, &
\Lambda_{2,1}^{(4)}(x)&=\bt 1,3 \\ 2,4 \et_x, &
\Lambda_{3,1}^{(4)}(x)&=\bt 1,2 \\ 2,3 \\ 3,4 \et_x,
\end{align}
although we should note that only $\Lambda_{1,1}^{(4)}(x)$ and $\Lambda_{3,1}^{(4)}(x)$ appear directly in our expressions. For convenience, in the calculations below we are going to consider the
normalized quantities
\begin{align}
\underline{\Lambda}_{1,1}^{(4)}(x)&=\Lambda_{1,1}^{(4)}(x), \\ 
\underline{\Lambda}_{3,1}^{(4)}(x)&=\frac{\Lambda_{3,1}^{(4)}(x)}{\Phi_-(x+\im)\Phi_+(x-\im)\Phi_-(x)\Phi_+(x)}.
\end{align}
Also, we define functions to replace the subgraphs appearing in \eqref{b4-11}--\eqref{b4-34} as follows from Sec.\ref{GRAPHS},
\begin{align}
\bt 1,2 \et_x
=&\frac{\Phi_+(x)p_{1,1}^{(4)}(x)}{q_2(x)},  &
\bt 2,3 \et_x
=&\frac{\Phi_-(x)\Phi_+(x)p_{1,2}^{(4)}(x)}{q_1(x)q_3(x)}, \nonumber\\
\bt 3,4 \et_x
=&\frac{\Phi_-(x)p_{1,3}^{(4)}(x)}{q_2(x)}, &
\bt 1,3 \et_x
=&\frac{\Phi_+(x)p_{1,4}^{(4)}(x)}{q_3(x)},\nonumber \\
\bt 2,4 \et_x
=&\frac{\Phi_-(x)p_{1,5}^{(4)}(x)}{q_1(x)}, \label{poly-sing}\\
\bt 1,2 \\ 2,3 \et_x
=&\frac{\Phi_+(x-\frac{\im}{2})\Phi_-(x+\frac{\im}{2})\Phi_+(x+\frac{\im}{2})p_{2,1}^{(4)}(x)}{q_3(x+\frac{\im}{2})}, & \bt 1,2 \\ 2,4 \et_x
=&\frac{\Phi_+(x-\frac{\im}{2})\Phi_-(x+\frac{\im}{2})p_{2,2}^{(4)}(x)}{q_2(x-\frac{\im}{2})}, \nonumber \\
\bt 1,3 \\ 3,4 \et_x
=&\frac{\Phi_+(x-\frac{\im}{2})\Phi_-(x+\frac{\im}{2})p_{2,3}^{(4)}(x)}{q_2(x+\frac{\im}{2})}, &
\bt 2,3 \\ 3,4 \et_x
=&\frac{\Phi_-(x-\frac{\im}{2})\Phi_+(x-\frac{\im}{2})\Phi_-(x+\frac{\im}{2})p_{2,4}^{(4)}(x)}{q_1(x-\frac{\im}{2})}.
\label{poly9}
\end{align}
By inspection, one finds that the zeros of all these {\unknown}s are located close to certain lines of the complex plane. In the case $\mu_j=0$, this is summarized in the table below.
\begin{center}
	\begin{tabular}{ |c|c| }
		\hline\hline
		\hbox{{\unknown}} & Im$(x)$ \\
		\hline\hline
		$q_j(x),\ j=1,\ldots,3$ & 0  \\
		\hline
		$p_{1,j}^{(4)}(x),\ j=1,\ldots,5$ & $\pm1$  \\
		\hline
		$p_{2,1}^{(4)}(x),\ p_{2,4}^{(4)}(x)$ & $\pm 3/2$  \\
		\hline
		$p_{2,2}^{(4)}(x)$ & $+1/2,\ \pm 3/2$ \\
		\hline
		$p_{2,3}^{(4)}(x)$ & $-1/2,\ \pm 3/2$ \\
		\hline
		$\underline{\Lambda}_{1,1}^{(4)}(x)$ & $\pm 1$ \\
		\hline
		$\underline{\Lambda}_{3,1}^{(4)}(x)$ & $\pm 3/2$ \\
		\hline
	\end{tabular}
\end{center}
It is worth noting that the zeros of $p_{1,j}^{(4)}(x)$ are known as the hole
solutions to the Bethe ansatz equations $\lambda_j/\lambda_{j+1}=-1$ for $j=1,2,3$. We want to obtain nontrivial relations among  $\{b_{a,j}^{(4)}(x)\}$ and
$\{B_{a,j}^{(4)}(x)\}$. To achieve this, the idea is to compute the
 logarithmic derivative of all functions and, in Fourier space,
express the {\unknown}s in
  terms of the $\{B_{a,j}^{(4)}(x)\}$. Inserting this into the expressions for
  the $\{b_{a,j}^{(4)}(x)\}$ and taking the inverse Fourier transform yields the NLIEs.
Thus we require these functions to be analytical, nonzero and with constant asymptotics (ANZ) in the strip $-\frac{1}{2}<\mathrm{Im}(x)<\frac{1}{2}$. Taking into account the analytical structure of the {\unknown}s according to the table above, we see that we must shift the argument of four functions as follows:
\begin{subequations}
\begin{align}
b_{2,1}^{(4)}(x-\im/2) \rightarrow b_{2,1}^{(4)}(x),&& B_{2,1}^{(4)}(x-\im/2) \rightarrow B_{2,1}^{(4)}(x),\\ b_{2,6}^{(4)}(x+\im/2) \rightarrow b_{2,6}^{(4)}(x), && b_{2,1}^{(4)}(x+\im/2) \rightarrow B_{2,6}^{(4)}(x), \label{shift1}
\end{align}
\end{subequations}
where we relabeled these functions as denoted. Then,
\begin{subequations}
	\begin{align}
	B_{1,1}^{(4)}(x)&=\frac{q_1(x+\frac{\im}{2})\underline{\Lambda}_{1,1}^{(4)}(x+\frac{\im}{2})}{\Phi_-(x+\frac{\im}{2})p_{1,5}^{(4)}(x+\frac{\im}{2})}, \label{b11}\\
	B_{1,2}^{(4)}(x)&= \frac{p_{1,1}^{(4)}(x-\frac{\im}{2})p_{1,5}^{(4)}(x+\frac{\im}{2})}{q_1(x+\frac{\im}{2})p_{2,2}^{(4)}(x)}, \\
	B_{1,3}^{(4)}(x)&= \frac{p_{1,3}^{(4)}(x+\frac{\im}{2})p_{1,4}^{(4)}(x-\frac{\im}{2})}{q_3(x-\frac{\im}{2})p_{2,3}^{(4)}(x)}, \\
	B_{1,4}^{(4)}(x)&= \frac{q_3(x-\frac{\im}{2})\underline{\Lambda}_{1,1}^{(4)}(x-\frac{\im}{2})}{\Phi_+(x-\frac{\im}{2})p_{1,4}^{(4)}(x-\frac{\im}{2})}, \\
	B_{2,1}^{(4)}(x)&= \frac{q_2(x+\frac{\im}{2})p_{2,2}^{(4)}(x)}{p_{1,1}^{(4)}(x-\frac{\im}{2})p_{1,3}^{(4)}(x+\frac{\im}{2})}, \\
	B_{2,2}^{(4)}(x)&= \frac{p_{1,2}^{(4)}(x)p_{2,3}^{(4)}(x+\frac{\im}{2})}{q_2(x+\im)p_{1,4}^{(4)}(x)p_{2,4}^{(4)}(x+\frac{\im}{2})}, \\
	B_{2,3}^{(4)}(x)&= \frac{p_{1,4}^{(4)}(x)p_{1,5}^{(4)}(x)}{p_{1,2}^{(4)}(x)\underline{\Lambda}_{1,1}^{(4)}(x)}, \\
	B_{2,4}^{(4)}(x)&= \frac{p_{2,1}^{(4)}(x-\frac{\im}{2})p_{2,4}^{(4)}(x+\frac{\im}{2})}{p_{1,2}^{(4)}(x)\underline{\Lambda}_{3,1}^{(4)}(x)}, \\
	B_{2,5}^{(4)}(x)&= \frac{p_{1,2}^{(4)}(x)p_{2,2}^{(4)}(x-\frac{\im}{2})}{q_2(x-\im)p_{1,5}^{(4)}(x)p_{2,1}^{(4)}(x-\frac{\im}{2})}, \\
	B_{2,6}^{(4)}(x)&= \frac{q_2(x-\frac{\im}{2})p_{2,3}^{(4)}(x)}{p_{1,1}^{(4)}(x-\frac{\im}{2})p_{1,3}^{(4)}(x+\frac{\im}{2})},\\
	B_{3,1}^{(4)}(x)&= \frac{q_3(x+\frac{3\im}{2})\underline{\Lambda}_{3,1}^{(4)}(x+\frac{\im}{2})}{\Phi_+(x+\frac{5\im}{2})p_{2,1}^{(4)}(x)}\ee^{-\beta\mu_4}, \\
	B_{3,2}^{(4)}(x)&= \frac{p_{1,3}^{(4)}(x+\frac{\im}{2})p_{2,1}^{(4)}(x)}{q_3(x+\frac{3\im}{2})p_{2,2}^{(4)}(x)}\ee^{-\beta\mu_3}, \\
	B_{3,3}^{(4)}(x)&= \frac{p_{1,1}^{(4)}(x-\frac{\im}{2})p_{2,4}^{(4)}(x)}{q_1(x-\frac{3\im}{2})p_{2,3}^{(4)}(x)}\ee^{-\beta\mu_2}, \\
	B_{3,4}^{(4)}(x)&= \frac{q_1(x-\frac{3\im}{2})\underline{\Lambda}_{3,1}^{(4)}(x-\frac{\im}{2})}{\Phi_-(x-\frac{5\im}{2})p_{2,4}^{(4)}(x)}\ee^{-\beta\mu_1},
	\end{align}
\end{subequations}
while their lowercase counterparts are
\begin{subequations}
\begin{align}
b_{1,1}^{(4)}(x)&= \frac{\Phi_-(x-\frac{\im}{2})\Phi_+(x+\frac{\im}{2})q_1(x+\frac{3\im}{2})}{\Phi_-(x+\frac{\im}{2})p_{1,5}^{(4)}(x+\frac{\im}{2})}\ee^{\beta\mu_1},  \\
b_{1,2}^{(4)}(x)&= \frac{\Phi_-(x-\frac{\im}{2})\Phi_+(x+\frac{\im}{2})q_1(x-\frac{3\im}{2})q_2(x+\frac{3\im}{2})}{q_1(x+\frac{\im}{2})p_{2,2}^{(4)}(x)}\ee^{2\beta\mu_2}, \\
b_{1,3}^{(4)}(x)&= \frac{\Phi_-(x-\frac{\im}{2})\Phi_+(x+\frac{\im}{2})q_2(x-\frac{3\im}{2})q_3(x+\frac{3\im}{2})}{q_3(x-\frac{\im}{2})p_{2,3}^{(4)}(x)}\ee^{2\beta\mu_3}, \\
b_{1,4}^{(4)}(x)&= \frac{\Phi_-(x-\frac{\im}{2})\Phi_+(x+\frac{\im}{2})q_3(x-\frac{3\im}{2})}{\Phi_+(x-\frac{\im}{2})p_{1,4}^{(4)}(x-\frac{\im}{2})}\ee^{\beta\mu_4},\\
b_{2,1}^{(4)}(x)&= \frac{\Phi_-(x-\frac{3\im}{2})\Phi_+(x+\frac{\im}{2})q_2(x-\frac{\im}{2})q_2(x+\frac{3\im}{2})}{p_{1,1}^{(4)}(x-\frac{\im}{2})p_{1,3}^{(4)}(x+\frac{\im}{2})}\ee^{\beta(\mu_1+\mu_2)}, \\
b_{2,2}^{(4)}(x)&= \frac{\Phi_-(x-\im)\Phi_+(x+\im)q_1(x+\im)q_2(x-\im)q_3(x+2\im)}{q_2(x+\im)p_{1,4}^{(4)}(x)p_{2,4}^{(4)}(x+\frac{\im}{2})}\ee^{\beta(\mu_1+2\mu_3)}, \\
b_{2,3}^{(4)}(x)&= \frac{\Phi_-(x-\im)\Phi_+(x+\im)q_1(x+\im)q_3(x-\im)}{p_{1,2}^{(4)}(x)\underline{\Lambda}_{1,1}^{(4)}(x)}\ee^{\beta(\mu_1+\mu_4)}, \\
b_{2,4}^{(4)}(x)&= \frac{\Phi_-(x-\im)\Phi_+(x+\im)q_1(x-2\im)q_3(x+2\im)}{p_{1,2}^{(4)}(x)\underline{\Lambda}_{3,1}^{(4)}(x)}\ee^{2\beta(\mu_2+\mu_3)}, \\
b_{2,5}^{(4)}(x)&= \frac{\Phi_-(x-\im)\Phi_+(x+\im)q_1(x-2\im)q_2(x+\im)q_3(x-\im)}{q_2(x-\im)p_{1,5}^{(4)}(x)p_{2,1}^{(4)}(x-\frac{\im}{2})}\ee^{\beta(2\mu_2+\mu_4)}, \\
b_{2,6}^{(4)}(x)&= \frac{\Phi_-(x-\frac{\im}{2})\Phi_+(x+\frac{3\im}{2})q_2(x+\frac{\im}{2})q_2(x-\frac{3\im}{2})}{p_{1,1}^{(4)}(x-\frac{\im}{2})p_{1,3}^{(4)}(x+\frac{\im}{2})}\ee^{\beta(\mu_3+\mu_4)}, \\
b_{3,1}^{(4)}(x)&= \frac{\Phi_-(x-\frac{3\im}{2})\Phi_+(x+\frac{3\im}{2})q_3(x+\frac{5\im}{2})}{\Phi_+(x+\frac{5\im}{2})p_{2,1}^{(4)}(x)}\ee^{\beta(\mu_1+\mu_2+\mu_3-\mu_4)}, \\
b_{3,2}^{(4)}(x)&= \frac{\Phi_-(x-\frac{3\im}{2})\Phi_+(x+\frac{3\im}{2})q_2(x+\frac{3\im}{2})q_3(x-\frac{\im}{2})}{q_3(x+\frac{3\im}{2})p_{2,2}^{(4)}(x)}\ee^{\beta(\mu_1+\mu_2-\mu_3+\mu_4)}, \\
b_{3,3}^{(4)}(x)&= \frac{\Phi_-(x-\frac{3\im}{2})\Phi_+(x+\frac{3\im}{2})q_1(x+\frac{\im}{2})q_2(x-\frac{3\im}{2})}{q_1(x-\frac{3\im}{2})p_{2,3}^{(4)}(x)}\ee^{\beta(\mu_1-\mu_2+\mu_3+\mu_4)}, \\
b_{3,4}^{(4)}(x)&= \frac{\Phi_-(x-\frac{3\im}{2})\Phi_+(x+\frac{3\im}{2})q_1(x-\frac{5\im}{2})}{\Phi_-(x-\frac{5\im}{2})p_{2,4}^{(4)}(x)}\ee^{\beta(-\mu_1+\mu_2+\mu_3+\mu_4)}. \label{bb34}
\end{align}
\end{subequations}
It is important to notice that when taking the Fourier transform of the logarithmic derivative of the auxiliary functions one must treat the cases $k<0$ and $k>0$
separately. After eliminating the unwanted variables, we transform the
resulting system back to real space and integrate with respect to $x$, which
leads us to a set of NLIEs involving only $\left\{b_{a,j}^{(4)}(x)\right\},
\left\{B_{a,j}^{(4)}(x)\right\}$, given by
\begin{align}
\log\mathbf{b}^{(4)}(x)=-\left(\mathbf{c}^{(4)}+\beta J\mathbf{d}^{(4)}(x)\right)-\mathbf{K}^{(4)}*\log\mathbf{B}^{(4)}(x), \label{nliesl4}
\end{align}
where $\mathbf{b}^{(4)}(x)=\left\{B_{a,j}^{(4)}(x)\right\}$ and
$\mathbf{B}^{(4)}(x)=\left\{B_{a,j}^{(4)}(x)\right\}$,
$a=1,\ldots,3,\ j=1,\ldots,d_a^{(4)}$. Here the logarithm is to be understood as applied to each
function in the set. Also, $\mathbf{c}$ is the constant of integration
$\mathbf{c}=\left\{c^{(4)}_1,\ldots,c^{(4)}_{14}\right\}$,
\begin{subequations}
\begin{align}
c^{(4)}_1&=\frac{\beta}{4} (-3 \mu_1+\mu_2+\mu_3+\mu_4),&&
c^{(4)}_2=\frac{\beta}{4} (\mu_1-3 \mu_2+\mu_3+\mu_4),\\
c^{(4)}_3&=\frac{\beta}{4} (\mu_1+\mu_2-3 \mu_3+\mu_4), &&
c^{(4)}_4=\frac{\beta}{4} (\mu_1+\mu_2+\mu_3-3 \mu_4),\\
c^{(4)}_5&=\frac{\beta}{2} (-\mu_1-\mu_2+\mu_3+\mu_4),&&
c^{(4)}_6=\frac{\beta}{2} (-\mu_1+\mu_2-\mu_3+\mu_4),\\
c^{(4)}_7&=\frac{\beta}{2} (-\mu_1+\mu_2+\mu_3-\mu_4),&&
c^{(4)}_8=\frac{\beta}{2}
(\mu_1-\mu_2-\mu_3+\mu_4),\\
c^{(4)}_9&=\frac{\beta}{2} (\mu_1-\mu_2+\mu_3-\mu_4),&&
c^{(4)}_{10}=\frac{\beta}{2} (\mu_1+\mu_2-\mu_3-\mu_4),\\
c^{(4)}_{11}&=\frac{\beta}{4} (-\mu_1-\mu_2-\mu_3+3
\mu_4),&&
c^{(4)}_{12}=\frac{\beta}{4} (-\mu_1-\mu_2+3 \mu_3-\mu_4),\\
c^{(4)}_{13}&=\frac{\beta}{4} (-\mu_1+3 \mu_2-\mu_3-\mu_4),&&
c^{(4)}_{14}=\frac{\beta}{4} (3 \mu_1-\mu_2-\mu_3-\mu_4),
\end{align}
\end{subequations}
which is obtained from the knowledge of the auxiliary functions as $x\rightarrow \infty$ together with the relation below for the convolution terms
\begin{align}
\lim\limits_{x\rightarrow\infty} f*g(x)
=g(\infty)\mathcal{F}[f(x)]_{k=0}, \label{conv-lim}
\end{align}
where $\mathcal{F}[f(x)]$ denotes the Fourier transform of the function $f(x)$. The driving term $\mathbf{d}^{(4)}(x)$ is
\begin{align}
\mathbf{d}^{(4)}(x)=\begin{bmatrix}
\mathbf{d}^{(4,1)}(x) \\ \mathbf{d}^{(4,2)}(x) \\ \mathbf{d}^{(4,3)}(x)
\end{bmatrix},\label{driv-su4}
\end{align}
where $\mathbf{d}^{(4,j)}(x)=\int_{-\infty}^{\infty}\ee^{\im k x} \hat{\mathbf{d}}^{(4,j)}(k)\dd k,\ j=1,2,3$, which in turn are given by

\begin{equation}
\hat{\mathbf{d}}^{(4,j)}(k)= \frac{\sinh\left( \left(\frac{4-j}{2}\right)k\right)}{\sinh(2 k)}\mathcal{T}_j^{-1} v_j,
\end{equation}
where $v_j$ are column vectors of dimension $d^{(4)}_j$ whose entries are 1, and $\mathcal{T}_j$ are diagonal matrices of this same dimension that encode the shifts of the auxiliary functions. As long as we do not violate ANZ strips, we can dispose of these shifts conveniently. For instance, for zero generalized chemical potentials we may assume that only $\mathcal{T}_2$ is different from identity and reads $\diag(y^{-1},1,1,1,1,y)$, with $y={\rm e}^{k/2}$.
$\mathbf{K}^{(4)}(x)$ is the kernel matrix, which can be written in terms of nine submatrices,
\begin{align}
\mathbf{K}^{(4)}(x)=\begin{bmatrix}
\mathbf{K}^{(4)}_{1,1}(x) & \mathbf{K}^{(4)}_{1,2}(x) & \mathbf{K}^{(4)}_{1,3}(x) \\
\mathbf{K}^{(4)}_{2,1}(x) & \mathbf{K}^{(4)}_{2,2}(x) & \mathbf{K}^{(4)}_{2,3}(x) \\
\mathbf{K}^{(4)}_{3,1}(x) & \mathbf{K}^{(4)}_{3,2}(x) & \mathbf{K}^{(4)}_{3,3}(x)
\end{bmatrix}.\label{ker-su4}
\end{align}
They satisfy the following symmetries:
\begin{align}
	\mathbf{K}^{(4)}(x) = \left[\mathbf{K}^{(4)}(x)\right]^{\dagger}, &&
\mathbf{K}^{(4)}_{i,j}(x)= {\cal T}_i^{-1} \Pi_{i} {\cal T}_{4-i}^{-1} {\left[\mathbf{K}^{(4) }_{4-j,4-i}(x)\right]}^t {\cal T}_{4-j} \Pi_{j} {\cal T}_{j},
\end{align}
where we have defined the reflection matrix $\left[\Pi_{j}\right]_{mn}= \delta_{d^{(4)}_j+1-m,n} $. Therefore only four blocks of them need to be explicitly written.

In Fourier space, these submatrices and their entries are given by
\begin{align}
\hat{\mathbf{K}}^{(4)}_{1,1}={\cal T}_{1}^{-1}\begin{bmatrix}
\hat{K}_0^{( 4)} & \hat{K}_1^{(4)} & \hat{K}_1^{(4)} & \hat{K}_1^{(4)} \\
\hat{K}_2^{(4)} & \hat{K}_0^{( 4)} & \hat{K}_1^{(4)} & \hat{K}_1^{(4)} \\
\hat{K}_2^{(4)} & \hat{K}_2^{(4)} & \hat{K}_0^{( 4)} & \hat{K}_1^{(4)} \\
\hat{K}_2^{(4)} & \hat{K}_2^{(4)} & \hat{K}_2^{(4)} & \hat{K}_0^{( 4)}
\end{bmatrix}{\cal T}_{1}
\end{align}
with
\begin{align}
\hat{K}_0^{( 4)}(k)&=\mathcal{K}_{[4]}^{(1,1)}(k),\\
\hat{K}_1^{(4)}(k)&=\mathcal{K}_{[4]}^{(1,1)}(k)+\ee^{-k/2-|k|/2},\\
\hat{K}_2^{(4)}(k)&=\mathcal{K}_{[4]}^{(1,1)}(k)+\ee^{k/2-|k|/2},
\end{align}
\begin{align}
\hat{\mathbf{K}}^{(4)}_{2,2}={\cal T}_{2}^{-1}\begin{bmatrix}
\hat{K}_3^{(4)} & \hat{K}_4^{(4)} & \hat{K}_4^{(4)} & \hat{K}_4^{(4)} & \hat{K}_4^{(4)} & \hat{K}_6^{(4)}  \\
\hat{K}_5^{(4)} & \hat{K}_3^{(4)} & \hat{K}_4^{(4)} & \hat{K}_4^{(4)} & \hat{K}_{8}^{(4)}  & \hat{K}_4^{(4)} \\
\hat{K}_5^{(4)} & \hat{K}_5^{(4)} & \hat{K}_3^{(4)} & \hat{K}_{10}^{(4)}  & \hat{K}_4^{(4)} & \hat{K}_4^{(4)} \\
\hat{K}_5^{(4)} & \hat{K}_5^{(4)} & \hat{K}_{10}^{(4)}  & \hat{K}_3^{(4)} & \hat{K}_4^{(4)} & \hat{K}_4^{(4)} \\
\hat{K}_5^{(4)} & \hat{K}_{9}^{(4)}  & \hat{K}_5^{(4)} & \hat{K}_5^{(4)} & \hat{K}_3^{(4)} & \hat{K}_4^{(4)} \\
\hat{K}_7^{(4)} & \hat{K}_5^{(4)} & \hat{K}_5^{(4)} & \hat{K}_5^{(4)} & \hat{K}_5^{(4)} &\hat{K}_3^{(4)}
\end{bmatrix} {\cal T}_{2},
\end{align}
where
\begin{align}
\hat{K}_3^{(4)}(k)&=\mathcal{K}_{[4]}^{(2,2)}(k),\\
\hat{K}_4^{(4)}(k)&=\mathcal{K}_{[4]}^{(2,2)}(k)+\ee^{-k/2-|k|/2},\\
\hat{K}_5^{(4)}(k)&=\mathcal{K}_{[4]}^{(2,2)}(k)+\ee^{k/2-|k|/2},\\
\hat{K}_6^{(4)} (k)&=\mathcal{K}_{[4]}^{(2,2)}(k)+\ee^{-k-|k|}-\ee^{-k},\label{k6}\\
\hat{K}_7^{(4)} (k)&=\mathcal{K}_{[4]}^{(2,2)}(k)+\ee^{k-|k|}-\ee^{k},\label{k7}\\
\hat{K}_{8}^{(4)} (k)&=\mathcal{K}_{[4]}^{(2,2)}(k)+2\ee^{-k/2-|k|/2},\\
\hat{K}_{9}^{(4)} (k)&=\mathcal{K}_{[4]}^{(2,2)}(k)+2\ee^{k/2-|k|/2},\\
\hat{K}_{10}^{(4)} (k)&=\mathcal{K}_{[4]}^{(2,2)}(k)+\ee^{-|k|}.
\end{align}
\begin{align}
\hat{\mathbf{K}}^{(4)}_{1,2}={\cal T}_{1}^{-1}\begin{bmatrix}
 \hat{K}_{11}^{(4)}  & \hat{K}_{11}^{(4)}  & \hat{K}_{11}^{(4)}  & \hat{K}_{12}^{(4)}  & \hat{K}_{12}^{(4)}  & \hat{K}_{12}^{(4)}  \\
 \hat{K}_{11}^{(4)}  & \hat{K}_{14}^{(4)}  & \hat{K}_{14}^{(4)}  & \hat{K}_{11}^{(4)}  & \hat{K}_{11}^{(4)}  & \hat{K}_{12}^{(4)}  \\
 \hat{K}_{13}^{(4)}  & \hat{K}_{11}^{(4)}  & \hat{K}_{14}^{(4)}  & \hat{K}_{11}^{(4)}  & \hat{K}_{14}^{(4)}  & \hat{K}_{11}^{(4)}  \\
 \hat{K}_{13}^{(4)}  & \hat{K}_{13}^{(4)}  & \hat{K}_{11}^{(4)}  & \hat{K}_{13}^{(4)}  & \hat{K}_{11}^{(4)}  & \hat{K}_{11}^{(4)}
\end{bmatrix}{\cal T}_{2},
\end{align}
with
\begin{align}
\hat{K}_{11}^{(4)} (k)&=\mathcal{K}_{[4]}^{(1,2)}(k),\\
\hat{K}_{12}^{(4)} (k)&=\mathcal{K}_{[4]}^{(1,2)}(k)+\ee^{-k-|k|/2}-\ee^{-k/2},\\
\hat{K}_{13}^{(4)} (k)&=\mathcal{K}_{[4]}^{(1,2)}(k)+\ee^{k-|k|/2}-\ee^{k/2},\\
\hat{K}_{14}^{(4)} (k)&=\mathcal{K}_{[4]}^{(1,2)}(k)+\ee^{-|k|/2},
\end{align}
and finally
\begin{align}
\hat{\mathbf{K}}^{(4)}_{1,3}={\cal T}_{1}^{-1}\begin{bmatrix}
\hat{K}_{15}^{(4)}  & \hat{K}_{15}^{(4)}  & \hat{K}_{15}^{(4)}  & \hat{K}_{16}^{(4)}  \\
\hat{K}_{15}^{(4)}  & \hat{K}_{15}^{(4)}  & \hat{K}_{18}^{(4)}  & \hat{K}_{15}^{(4)}  \\
\hat{K}_{15}^{(4)}  & \hat{K}_{19}^{(4)}  & \hat{K}_{15}^{(4)}  & \hat{K}_{15}^{(4)}  \\
\hat{K}_{17}^{(4)}  & \hat{K}_{15}^{(4)}  & \hat{K}_{15}^{(4)}  & \hat{K}_{15}^{(4)}
\end{bmatrix}{\cal T}_{3},
\end{align}
with
\begin{align}
\hat{K}_{15}^{(4)} (k)&=\mathcal{K}_{[4]}^{(1,3)}(k),\\
\hat{K}_{16}^{(4)} (k)&=\mathcal{K}_{[4]}^{(1,3)}(k)+\ee^{-3k/2-|k|/2}-\ee^{-k},\\
\hat{K}_{17}^{(4)} (k)&=\mathcal{K}_{[4]}^{(1,3)}(k)+\ee^{3k/2-|k|/2}-\ee^{k},\\
\hat{K}_{18}^{(4)} (k)&=\mathcal{K}_{[4]}^{(1,3)}(k)+\ee^{-k/2-|k|/2},\label{k18}\\
\hat{K}_{19}^{(4)} (k)&=\mathcal{K}_{[4]}^{(1,3)}(k)+\ee^{k/2-|k|/2},
\end{align}
where the function $\mathcal{K}_{[4]}^{(a,b)}(k)$ appearing everywhere is
defined by
\begin{align}
\mathcal{K}^{(a,b)}_{[n]}(k)=\ee^{|k|/2}\frac{\sh\left[\min\left(a,b\right)\frac{k}{2}\right]\sh\left[\left(n-\max\left(a,b\right)\right)\frac{k}{2}\right]}{\sh\left(\frac{k}{2}\right)\sh\left(\frac{nk}{2}\right)}-\delta_{a,b}. \label{common}
\end{align}
The NLIEs (\ref{nliesl4}) allow to calculate the functions
$\{B_{a,j}^{(4)}=1+b_{a,j}^{(4)}\}$ by for instance numerical iterations.

Now, computing the largest eigenvalue of the QTM
$\Lambda_{1,1}^{(4)}(x)$ is relatively simple. Indeed, one of the expressions
of the {\unknown}s in terms of the $\{B_{a,j}^{(4)}\}$ yields
\begin{align}
\lim_{N \rightarrow \infty}\log \frac{\Lambda_{1,1}^{(4)}(x)}{\Phi_+(x-\im) \Phi_-(x+\im)} = -\im \beta J\partial_x \log \frac{\Gamma(
	\frac{1}{4}- \im \frac{x}{4}) \Gamma(1 + \im \frac{x}{4})}{\Gamma(\frac{1}{4} + \im \frac{x}{4}) \Gamma(1 - \im \frac{x}{4})}+ \beta \sum_{j=1}^{4}\frac{\mu_j}{4} + \mathbf{d}^{(4)}(x)^{\dagger} \ast \log \mathbf{B}^{(4)}(x),
\end{align}
with $\mathbf{d}^{(4)}(x)$ given by \eqref{driv-su4}.

It is noteworthy that the above results are much alike those of
\cite{DAMERAU}. We recall that the authors have found by trial and error the
same number of auxiliary functions as one does by following our method and
actually many of them coincide with ours. Furthermore the
  functions that are distinct
are related to each other by means of the function $f^{(4)}(x)$,
\begin{align}
f^{(4)}(x)&=\frac{\bt 1,2\\2,4 \et_x \bt 1,3 \\ 3,4 \et_x}{\bt 1,3 \\ 2,4 \et_x \bt 1,2 \\ 3,4 \et_x }=\frac{\Phi_-\left(x+\frac{\im}{2}\right)\Phi_+\left(x-\frac{\im}{2}\right)p_{2,2}^{(4)}(x) p_{2,3}^{(4)}(x) }{p_{1,1}^{(4)}\left(x-\frac{\im}{2}\right)p_{1,3}^{(4)} \left(x+\frac{\im}{2}\right)\Lambda_{2,1}^{(4)}(x)},
\end{align}
as
\begin{align}
\mathsf{B}_{1,2}^{(4)}(x)&=f^{(4)}(x)B_{1,2}^{(4)}(x), && \mathsf{B}_{3,3}^{(4)}(x)=f^{(4)}(x)B_{3,3}^{(4)}(x) \\
\mathsf{B}_{2,1}^{(4)}(x)&=\frac{1}{f^{(4)}\left(x-\frac{\im}{2}\right)}B_{2,1}^{(4)}(x), && \mathsf{B}_{2,6}^{(4)}(x)=\frac{1}{f^{(4)}\left(x+\frac{\im}{2}\right)}B_{2,6}^{(4)}(x),
\end{align}
where $\mathsf{B}_{1,2}^{(4)}(x),\ \mathsf{B}_{2,1}^{(4)}(x),\ \mathsf{B}_{2,6}^{(4)}(x)\ \mathsf{B}_{3,3}^{(4)}(x)$ are given in Eqs.\eqref{jens-bb1}--\eqref{jens-bb14}. Note that $f^{(4)}(x)$ is self-conjugated. When compared with the Y-system, 
we obtain
\begin{align}
Y_{1}^{(4)}(x)&=f^{(4)}(x)\prod_{j=1}^{4}B^{(4)}_{1,j}(x),\\
Y_{2}^{(4)}(x)&=\frac{1}{f^{(4)}\left(x-\frac{\im}{2}\right)f^{(4)}\left(x+\frac{\im}{2}\right)}\prod_{j=1}^{6}B^{(4)}_{2,j}(x),\\
Y_{3}^{(4)}(x)&=f^{(4)}(x)\prod_{j=1}^{4}B^{(4)}_{3,j}(x).
\end{align}
Hence, the products of our uppercase functions within each representation do not give the Y-system \textit{per se} but are related to them via $f^{(4)}(x)$ in a simple way.

As a final remark, regarding the NLIEs we note that if one disregards the
shifts \eqref{shift1}, whose effect is to introduce the factors $y=\ee^{k/2}$
(or $1/y$) our driving term and kernel matrices look the same and are given by
the same functions as in \cite{DAMERAU} except for \eqref{k6}, \eqref{k7} and
\eqref{k18} which differ by $\ee^{-k},\ \ee^{k}$ and 1, respectively.

\subsubsection{$sl_5$}
\normalsize
In the $sl_5$ case the canonical construction gives us 30 lower case and 30 upper case
functions. The first and last representations are conjugated to one another as well as the second and third. For $a=1$ we have
\begin{subequations}
\begin{align}
B_{1,1}^{(5)}(x)&=\frac{\bt 1,5 \et_{x+\im/2}}{\bt 2,5 \et_{x+\im/2}}, & b_{1,1}^{(5)}(x)&=\frac{\bt 1 \et_{x+\im/2}}{\bt 2,5 \et_{x+\im/2}},  \\
B_{1,2}^{(5)}(x)&=\frac{\bt 1,2 \et_{x-\im/2} \bt 2,5 \et _{x+\im/2}}{\bt 1,2 \\ 2,5 \et_x}, & b_{1,2}^{(5)}(x)&=\frac{\bt 2 \et_{x-\im/2} \bt 2 \et_{x+\im/2}}{\bt 1,2 \\ 2,5 \et_x}, \\
B_{1,3}^{(5)}(x)&=\frac{\bt 1, 3 \et_{x-\im/2} \bt 3,5 \et _{x+\im/2}}{\bt 1,3 \\ 3,5 \et_x}, & b_{1,3}^{(5)}(x)&=\frac{\bt 3  \et_{x-\im/2} \bt 3 \et_{x+\im/2}}{\bt 1,3 \\ 3,5 \et_x}, \\
B_{1,4}^{(5)}(x)&=\frac{\bt 1,4 \et_{x-\im/2} \bt 4,5 \et_{x+\im/2}}{\bt 1,4 \\ 4,5 \et_x}, & b_{1,4}^{(5)}(x)&=\frac{\bt 4 \et_{x-\im/2} \bt 4 \et_{x-\im/2}}{\bt 1,4 \\ 4,5 \et_x},  \\
B_{1,5}^{(5)}(x)&=\frac{\bt 1,5 \et_{x-\im/2}}{\bt 1, 4 \et_{x-\im/2}}, & b_{1,5}^{(5)}(x)&=\frac{\bt 5 \et_{x-\im/2}}{\bt 1,4 \et_{x-\im/2}},
\end{align}
\end{subequations}
then, in the $a=2$ representation,
\begin{subequations}
\begin{align}
B_{2,1}^{(5)}(x)&=\frac{\bt 1,2 \\ 2,5 \et_{x+\im/2}}{\bt 1,2 \et_{x} \bt 3, 5 \et_{x+\im}}, & b_{2,1}^{(5)}(x)&=\frac{\bt 1 \\ 2 \et_{x+\im/2}}{\bt 1,2 \et_x \bt 3,5 \et_{x+\im}}, \\
B_{2,2}^{(5)}(x)&=\frac{\bt 2,3 \et_x \bt 1,3 \\ 3,5 \et_{x+\im/2}}{\bt 1,3 \et_x \bt 2,3 \\ 3,5 \et_{x+\im/2}}, & b_{2,2}^{(5)}(x)&=\frac{\bt 3 \et_x \bt 1 \\ 3 \et_{x+\im/2}}{\bt 1,3 \et_x \bt 2,3 \\ 3,5 \et_{x+\im/2}}, \\
B_{2,3}^{(5)}(x)&=\frac{\bt 1,2 \\ 2,3 \et_{x-\im/2} \bt 2,3 \\ 3,5 \et_{x+\im/2}}{\bt 2,3 \et_x \bt 1,2 \\ 2,3 \\ 3,5 \et_x}, & b_{2,3}^{(5)}(x)&=\frac{\bt 2 \\ 3 \et_{x-\im/2} \bt 2 \\ 3 \et_{x+\im/2}}{\bt 2,3 \et_x \bt 1,2 \\ 2,3 \\ 3,5 \et_x}\\
B_{2,4}^{(5)}(x)&=\frac{\bt 2,4 \et_x \bt 1,4 \\ 4,5 \et_{x+\im/2}}{\bt 1,4 \et_x \bt 2,4 \\ 4,5 \et_{x+\im/2}}, & b_{2,4}^{(5)}(x)&=\frac{\bt 4 \et_x \bt 1 \\ 4 \et_{x+\im/2}}{\bt 1,4 \et_x \bt 2,4 \\ 4,5 \et_{x+\im/2}}, \\
B_{2,5}^{(5)}(x)&=\frac{\bt 1,4 \et_x \bt 2,5 \et_x}{\bt 1, 5 \et_x \bt 2,4 \et_x}, &
b_{2,5}^{(5)}(x)&=\frac{\bt 1 \et_x  \bt 5 \et_x}{\bt 1, 5 \et_x \bt 2,4 \et_x},\\
B_{2,6}^{(5)}(x)&=\frac{\bt 1,2 \\ 2,4 \et_{x-\im/2} \bt 2,4 \\ 4,5 \et_{x+\im/2}}{\bt 2,4 \et_x \bt 1,2 \\ 2,4 \\ 4,5 \et_x}, & b_{2,6}^{(5)}(x)&=\frac{\bt 2 \\ 4 \et_{x-\im/2} \bt 2 \\ 4 \et_{x+\im/2}}{\bt 2,4 \et_x \bt 1,2 \\ 2,4 \\ 4,5 \et_x}\\
B_{2,7}^{(5)}(x)&=\frac{\bt 2,4 \et_x \bt 1,2 \\ 2,5 \et_{x-\im/2} }{\bt 2,5 \et_x \bt 1,2 \\ 2,4 \et_{x-\im/2} }, & b_{2,7}^{(5)}(x)&=\frac{\bt 2 \et_x \bt 2 \\ 5 \et_{x-\im/2}}{\bt 2,5 \et_x \bt 1,2 \\ 2,4 \et_{x-\im/2} }, \\
B_{2,8}^{(5)}(x)&=\frac{\bt 1,3 \\ 3,4 \et_{x-\im/2} \bt 3,4 \\ 4,5 \et_{x+\im/2}}{\bt 3,4 \et_x \bt 1,3 \\ 3,4 \\ 4,5 \et_x}, & b_{2,8}^{(5)}(x)&=\frac{\bt 3 \\ 4 \et_{x-\im/2} \bt 3 \\ 4 \et_{x+\im/2}}{\bt 3,4 \et_x \bt 1,3 \\ 3,4 \\ 4,5 \et_x}, \\
B_{2,9}^{(5)}(x)&=\frac{\bt 3,4 \et_x \bt 1,3 \\ 3,5 \et_{x-\im/2}}{\bt 3, 5 \et_x \bt 1,3 \\ 3,4 \et_{x-\im/2}}, & b_{2,9}^{(5)}(x)&=\frac{\bt 3 \et_x \bt 3 \\ 5 \et_{x-\im/2}}{\bt 3, 5 \et_x \bt 1,3 \\ 3,4 \et_{x-\im/2}}, \\
B_{2,10}^{(5)}(x)&=\frac{\bt 1,4 \\ 4,5 \et_{x-\im/2}}{\bt 1,3 \et_{x-\im} \bt 4, 5 \et_x}, & b_{2,10}^{(5)}(x)&=\frac{\bt 4 \\ 5 \et_{x-\im/2}}{\bt 1,3 \et_{x-\im} \bt 4, 5 \et_x},
\end{align}
\end{subequations}
conjugated to the functions in the $a=3$ representation, which in turn are given by
\begin{subequations}
\begin{align}
B_{3,1}^{(5)}(x)&=\frac{\bt 1,2 \\ 2,3 \\ 3,5 \et_{x+\im/2}}{\bt 4,5 \et_{x+3\im/2}  \bt 1,2 \\ 2,3 \et_{x} }, & b_{3,1}^{(5)}(x)&=\frac{ \bt 1 \\ 2 \\ 3 \et_{x+\im/2}}{\bt 4,5 \et_{x+3\im/2}  \bt 1,2 \\ 2,3 \et_{x} }, \\
B_{3,2}^{(5)}(x)&=\frac{\bt 3,4 \et_{x+\im/2} \bt 1,2 \\ 2,4 \\ 4,5 \et_{x+\im/2}}{\bt 1,2 \\ 2,4 \et_{x} \bt 3,4 \\ 4,5 \et_{x+\im}}, & b_{3,2}^{(5)}(x)&=\frac{\bt 1 \\ 4 \et_x \bt 2 \\ 4 \et_{x+\im}}{\bt 1,2 \\ 2,4 \et_{x} \bt 3,4 \\ 4,5 \et_{x+\im}},\\
B_{3,3}^{(5)}(x)&=\frac{\bt 3,5 \et_{x+\im/2} \bt 1,2 \\ 2,4 \et_x }{\bt 3,4 \et_{x+\im/2} \bt 1,2 \\ 2,5 \et_x }, & b_{3,3}^{(5)}(x)&=\frac{\bt 5 \et_{x+\im/2} \bt 1 \\ 2 \et_x}{\bt 3,4 \et_{x+\im/2} \bt 1,2 \\ 2,5 \et_x },\\
B_{3,4}^{(5)}(x)&=\frac{\bt 2,3 \\ 3,4 \et_x \bt 1,3 \\ 3,4 \\ 4,5 \et_{x+\im/2}}{\bt 1,3 \\ 3,4 \et_x \bt 2,3 \\ 3,4 \\ 4,5 \et_{x+\im/2}}, & b_{3,4}^{(5)}(x)&=\frac{\bt 3 \\ 4 \et_x \bt 1 \\ 3 \\ 4 \et_{x+\im/2}}{\bt 1,3 \\ 3,4 \et_x \bt 2,3 \\ 3,4 \\ 4,5 \et_{x+\im/2}} \\
B_{3,5}^{(5)}(x)&=\frac{\bt 1,3 \\ 3,4 \et_x \bt 2,3 \\ 3,5 \et _x}{\bt 2,3 \\ 3,4 \et_x \bt 1,3 \\ 3,5 \et_x}, & b_{3,5}^{(5)}(x)&=\frac{\bt 1 \\ 3 \et_x \bt 3 \\ 5 \et_x }{\bt 2,3 \\ 3,4 \et_x \bt 1,3 \\ 3,5 \et_x},\\
B_{3,6}^{(5)}(x)&=\frac{\bt 1,2 \\ 2,3 \\ 3,4 \et_{x-\im/2} \bt 2,3 \\ 3,4 \\ 4,5 \et_{x+\im/2}}{\bt 2,3 \\ 3,4 \et_x \bt 1,2 \\ 2,3 \\ 3,4 \\ 4,5 \et_x}, & b_{3,6}^{(5)}(x)&=\frac{\bt 2 \\ 3 \\ 4 \et_{x-\im/2}  \bt 2 \\ 3 \\ 4 \et_{x+\im/2} }{\bt 2,3 \\ 3,4 \et_x \bt 1,2 \\ 2,3 \\ 3,4 \\ 4,5 \et_x},\\
B_{3,7}^{(5)}(x)&=\frac{\bt 2,3 \\ 3,4 \et_x \bt 1,2 \\ 2,3 \\ 3,5 \et_{x-\im/2} }{\bt 2,3 \\ 3,5 \et_x \bt 1,2 \\ 2,3 \\ 3,4 \et_{x-\im/2} }, & b_{3,7}^{(5)}(x)&=\frac{\bt 2 \\ 3 \et_x \bt 2 \\ 3 \\ 5 \et_{x-\im/2}}{\bt 2,3 \\ 3,5 \et_x \bt 1,2 \\ 2,3 \\ 3,4 \et_{x-\im/2} },\\
B_{3,8}^{(5)}(x)&=\frac{\bt 1,3 \et_{x-\im/2} \bt 2,4 \\ 4,5 \et_x}{\bt 2,3 \et_{x-\im/2} \bt 1,4 \\ 4,5 \et_x}, & b_{3,8}^{(5)}(x)&=\frac{\bt 1 \et_{x-\im/2} \bt 4 \\ 5 \et_x}{\bt 2,3 \et_{x-\im/2} \bt 1,4 \\ 4,5 \et_x},\\
B_{3,9}^{(5)}(x)&=\frac{\bt 2,3 \et_{x-\im/2} \bt 1,2 \\ 2,4 \\ 4,5 \et_{x-\im/2}}{\bt 1,2 \\ 2,3 \et_{x-\im} \bt 2,4 \\ 4,5 \et_x}, & b_{3,9}^{(5)}(x)&=\frac{\bt 2 \\ 4 \et_{x-\im} \bt 2 \\ 5 \et_x}{\bt 1,2 \\ 2,3 \et_{x-\im} \bt 2,4 \\ 4,5 \et_x},\\
B_{3,10}^{(5)}(x)&=\frac{\bt 1,3 \\ 3,4 \\ 4,5 \et_{x-\im/2}}{\bt 1,2 \et_{x-3\im/2} \bt 3,4 \\ 4,5 \et_x}, & b_{3,10}^{(5)}(x)&=\frac{ \bt 3 \\ 4 \\ 5 \et_{x-\im/2}}{\bt 1,2 \et_{x-3\im/2} \bt 3,4 \\ 4,5 \et_x} ,
\end{align}
\end{subequations}
and, finally, in the $a=4$ representation,
\begin{subequations}
\begin{align}
B_{4,1}^{(5)}(x)&=\frac{\bt 1,2 \\ 2,3 \\ 3,4 \\ 4,5 \et_{x+\im/2}}{\bt 1,2 \\ 2,3 \\ 3,4 \\ 5 \et_{x+\im/2}}, & b_{4,1}^{(5)}(x)&=\frac{\bt 1 \\ 2 \\ 3 \\ 4 \et_{x+\im/2}}{\bt 1,2 \\ 2,3 \\ 3,4 \\ 5 \et_{x+\im/2}},\\
B_{4,2}^{(5)}(x)&=\frac{\bt 4,5 \et_{x+\im} \bt 1,2 \\ 2,3 \\ 3,4 \et_x }{\bt 4 \et_{x+\im} \bt 1,2 \\ 2,3 \\ 3,5 \et_x }, & b_{4,2}^{(5)}(x)&=\frac{\bt 5 \et_{x+\im} \bt 1 \\ 2 \\ 3 \et_x}{\bt 4 \et_{x+\im} \bt 1,2 \\ 2,3 \\ 3,5 \et_x} ,\\
B_{4,3}^{(5)}(x)&=\frac{\bt 1,2 \\ 2,3 \et_{x-\im/2} \bt 3,4 \\ 4,5 \et_{x+\im/2}}{\bt 3 \et_x \bt 1,2 \\ 2,4 \\ 4,5 \et_x}, & b_{4,3}^{(5)}(x)&=\frac{\bt 2 \et_x \bt 1 \\ 4 \\ 5 \et_x}{\bt 3 \et_x \bt 1,2 \\ 2,4 \\ 4,5 \et_x}, \\
B_{4,4}^{(5)}(x)&=\frac{\bt 1,2 \et_{x-\im} \bt 2,3 \\ 3,4 \\ 4,5 \et_x}{\bt 2 \et_{x-\im} \bt 1,3 \\ 3,4 \\ 4,5 \et_x}, & b_{4,4}^{(5)}(x)&=\frac{\bt 1 \et_{x-\im} \bt 3 \\ 4 \\ 5 \et_x}{\bt 2 \et_{x-\im} \bt 1,3 \\ 3,4 \\ 4,5 \et_x}, \\
B_{4,5}^{(5)}(x)&=\frac{\bt 1,2 \\ 2,3 \\ 3,4 \\ 4,5 \et_{x-\im/2}}{\bt 1 \\ 2,3 \\ 3,4 \\ 4,5 \et_{x-\im/2}}, & b_{4,5}^{(5)}(x)&=\frac{\bt 2 \\ 3 \\ 4 \\ 5 \et_{x-\im/2}}{\bt 1 \\ 2,3 \\ 3,4 \\ 4,5 \et_{x-\im/2}}.
\end{align}
\end{subequations}
It is straightforward to show that these auxiliary functions relate to the
Y-system through a common function, $f^{\lp 5 \rp}\lp x \rp$,
\begin{align}
	f^{(5)}(x)= \frac{\bt 1,2 \\ 2,5 \et_x \bt 1,3 \\ 3,5 \et_x \bt 1,4 \\ 4,5 \et_x}{\bt 1,4 \\ 2,5 \et_x \bt 1,2 \\ 3,5 \et_x \bt 1,3 \\ 4,5 \et_x}, \label{f5}
\end{align}
as follows:
\begin{align}
	Y^{(5)}_1(x) &= f^{(5)}(x) \prod_{j=1}^{5} B_{1,j}^{(5)}(x), \\
	Y^{(5)}_2(x) &= \frac{\overline{f^{(5)}}(x)}{f^{(5)}\left(x+\frac{\im}{2}\right)f^{(5)}\left(x-\frac{\im}{2}\right)} \prod_{j=1}^{10} B_{2,j}^{(5)}(x), \\
	Y^{(5)}_3(x) &= \frac{f^{(5)}(x)}{\overline{f^{(5)}}\left(x+\frac{\im}{2}\right)\overline{f^{(5)}}\left(x-\frac{\im}{2}\right)} \prod_{j=1}^{10} B_{3,j}^{(5)}(x), \\
	Y^{(5)}_4(x) &= \overline{f^{(5)}}(x) \prod_{j=1}^{5} B_{4,j}^{(5)}(x),
\end{align}
where the bar denotes representation conjugation. It is worth
noting that while the multiplication of function $f^{(4)}(x)$ by some
auxiliary functions lead to a perfect factorization of the Y-system in the
$sl_4$ case (i.e.~with the modified auxiliary functions the product within
the $a$-th representations lead to $Y_a^{(4)}(x)$), one cannot redistribute all
the factors appearing on Eq.\eqref{f5} among the canonical auxiliary functions
$B_{a,j}^{(5)}(x)$ and obtain another suitable set of functions (i.e.~that
preserves the relation $B(x)=1+b(x)$) which features this property.

In order to obtain a system of NLIEs for the auxiliary functions in this case
we apply the same reasoning as the previous section. Because of the
analyticity properties of the {\unknown}s we must shift some of the functions
(and relabel them) as follows:
\begin{subequations}
\begin{align}
b_{1,j}^{(5)}(x+\im/4) \rightarrow b_{1,j}^{(5)}(x),&& B_{1,j}^{(5)}(x+\im/4) \rightarrow B_{1,j}^{(5)}(x), && j=4,5,\\
b_{2,j}^{(5)}(x-\im/2) \rightarrow b_{2,j}^{(5)}(x),&& B_{2,j}^{(5)}(x-\im/2) \rightarrow B_{2,j}^{(5)}(x), && j=1,2,3,\\
b_{2,j}^{(5)}(x+\im/2) \rightarrow b_{2,j}^{(5)}(x),&& B_{2,j}^{(5)}(x+\im/2) \rightarrow B_{2,j}^{(5)}(x), && j=8,9,\\
b_{2,10}^{(5)}(x+3\im/4) \rightarrow b_{2,10}^{(5)}(x),&& B_{2,10}^{(5)}(x+3\im/4) \rightarrow B_{2,10}^{(5)}(x), &&\\
b_{3,1}^{(5)}(x-3\im/4) \rightarrow b_{3,1}^{(5)}(x),&& B_{3,1}^{(5)}(x-3\im/4) \rightarrow B_{3,1}^{(5)}(x), &&  \\
b_{3,10}^{(5)}(x+\im) \rightarrow b_{3,10}^{(5)}(x),&& B_{3,10}^{(5)}(x+\im) \rightarrow B_{3,10}^{(5)}(x), && \\
b_{4,j}^{(5)}(x-\im/2) \rightarrow b_{4,j}^{(5)}(x),&& B_{4,j}^{(5)}(x-\im/2) \rightarrow B_{4,j}^{(5)}(x), && j=1,2,\\
b_{4,j}^{(5)}(x+\im/2) \rightarrow b_{4,j}^{(5)}(x),&& B_{4,j}^{(5)}(x+\im/2) \rightarrow B_{4,j}^{(5)}(x), && j=4,5.
\label{shift2}
\end{align}
\end{subequations}
From here on the steps are the same as explained in the $sl_4$ case so we
refrain from repeating them and simply state the results. The NLIEs can be
written in the same form as Eq.\eqref{nliesl4},
\begin{align}
\log\mathbf{b}^{(5)}(x)=-\left(\mathbf{c}^{(5)}+\beta J\mathbf{d}^{(5)}(x)\right)-\mathbf{K}^{(5)}*\log\mathbf{B}^{(5)}(x),
\end{align}
with the vectors defined similarly as before. The integration constants are given by
\begin{subequations}
	\begin{align}
	c^{(5)}_ 1&=\frac{\beta}{5} (-4 \mu_1+\mu_2+\mu_3+\mu_4+\mu_5),&&
	c^{(5)}_ 2=\frac{\beta}{5} (\mu_1-4 \mu_2+\mu_3+\mu_4+\mu_5),\\
	c^{(5)}_ 3&=\frac{\beta}{5} (\mu_1+\mu_2-4 \mu_3+\mu_4+\mu_5),&&
	c^{(5)}_ 4=\frac{\beta}{5} (\mu_1+\mu_2+\mu_3-4 \mu_4+\mu_5),\\
	c^{(5)}_ 5&=\frac{\beta}{5} (\mu_1+\mu_2+\mu_3+\mu_4-4 \mu_5),&&
	c^{(5)}_ 6=\frac{\beta}{5} (-3 \mu_1-3 \mu_2+2 \mu_3+2\mu_4+2\mu_5),\\
	c^{(5)}_ 7&=\frac{\beta}{5} (-3 \mu_1+2 \mu_2-3 \mu_3+2 \mu_4+2 \mu_5),&&
	c^{(5)}_ 8=\frac{\beta}{5} (2 \mu_1-3
	\mu_2-3 \mu_3+2 \mu_4+2 \mu_5),\\
	c^{(5)}_ 9&=\frac{\beta}{5} (-3 \mu_1+2 \mu_2+2 \mu_3-3 \mu_4+2 \mu_5),&&
	c^{(5)}_ {10}=\frac{\beta}{5} (-3 \mu_1+2 \mu_2+2 \mu_3+2 \mu_4-3 \mu_5),\\
	c^{(5)}_ {11}&=\frac{\beta}{5} (2
	\mu_1-3 \mu_2+2 \mu_3-3 \mu_4+2 \mu_5),&&
	c^{(5)}_ {12}=\frac{\beta}{5} (2 \mu_1-3 \mu_2+2 \mu_3+2 \mu_4-3 \mu_5),\\
	c^{(5)}_ {13}&=\frac{\beta}{5} (2 \mu_1+2 \mu_2-3 \mu_3-3 \mu_4+2 \mu_5),&&
	c^{(5)}_ {14}=\frac{\beta}{5} (2 \mu_1+2 \mu_2-3 \mu_3+2 \mu_4-3 \mu_5),\\
	c^{(5)}_ {15}&=\frac{\beta}{5} (2 \mu_1+2
	\mu_2+2 \mu_3-3 \mu_4-3 \mu_5),&&
	c^{(5)}_ {16}=\frac{\beta}{5} (-2 \mu_1-2 \mu_2-2 \mu_3+3 \mu_4+3 \mu_5),\\
	c^{(5)}_ {17}&=\frac{\beta}{5} (-2 \mu_1-2 \mu_2+3 \mu_3-2 \mu_4+3 \mu_5),&&
	c^{(5)}_ {18}=\frac{\beta}{5} (-2
	\mu_1-2 \mu_2+3 \mu_3+3 \mu_4-2 \mu_5),\\
	c^{(5)}_ {19}&=\frac{\beta}{5} (-2 \mu_1+3 \mu_2-2 \mu_3-2 \mu_4+3 \mu_5),&&
	c^{(5)}_ {20}=\frac{\beta}{5} (-2 \mu_1+3 \mu_2-2 \mu_3+3 \mu_4-2 \mu_5),\\
	c^{(5)}_ {21}&=\frac{\beta}{5} (3 \mu_1-2 \mu_2-2 \mu_3-2 \mu_4+3 \mu_5),&&
	c^{(5)}_ {22}=\frac{\beta}{5} (3 \mu_1-2
	\mu_2-2 \mu_3+3 \mu_4-2 \mu_5),\\
	c^{(5)}_ {23}&=\frac{\beta}{5} (-2 \mu_1+3 \mu_2+3 \mu_3-2 \mu_4-2 \mu_5),&&
	c^{(5)}_ {24}=\frac{\beta}{5} (3 \mu_1-2 \mu_2+3 \mu_3-2 \mu_4-2 \mu_5),\\
	c^{(5)}_ {25}&=\frac{\beta}{5} (3
	\mu_1+3 \mu_2-2 \mu_3+2\mu_4+2\mu_5),&&
	c^{(5)}_ {26}=\frac{\beta}{5} (-\mu_1-\mu_2-\mu_3-\mu_4+4 \mu_5),\\
	c^{(5)}_ {27}&=\frac{\beta}{5} (-\mu_1-\mu_2-\mu_3+4 \mu_4-\mu_5),&&
	c^{(5)}_ {28}=\frac{\beta}{5}
	(-\mu_1-\mu_2+4 \mu_3-\mu_4-\mu_5),\\
	c^{(5)}_ {29}&=\frac{\beta}{5} (-\mu_1+4 \mu_2-\mu_3-\mu_4-\mu_5),&&
	c^{(5)}_ {30}=\frac{\beta}{5} (4 \mu_1-\mu_2-\mu_3-\mu_4-\mu_5).
	\end{align}
\end{subequations}
On the other hand, the driving term $\mathbf{d}^{(5)}(x)$ is
\begin{align}
\mathbf{d}^{(5)}(x)=\begin{bmatrix}
\mathbf{d}^{(5,1)}(k) \\ \mathbf{d}^{(5,2)}(k) \\ \mathbf{d}^{(5,3)}(k) \\
\mathbf{d}^{(5,4)}(k)
\end{bmatrix},\label{driv-su5}
\end{align}
where the entries of this vector are the inverse Fourier transforms of
\begin{equation}
\hat{\mathbf{d}}^{(5,j)}(k)= \frac{\sinh\left( (5-j)k/2\right)}{\sinh(5 k/2)}\mathcal{T}_j^{-1} v_j,
\end{equation}
where $v_j$ are column vectors of dimension $d^{(5)}_j$ whose entries are 1, and $\mathcal{T}_j$ are diagonal matrices of this same dimension, that encode the shifts of the auxiliary functions. From the above, one can read that our preferred choice was $\mathcal{T}_1=\diag(1,1,1,y^{1/2},y^{1/2}),~\mathcal{T}_2=\diag(y^{-1},y^{-1},y^{-1},1,1,1,1,y,y,y^{3/2}),~\mathcal{T}_3=\diag(y^{-3/2},1,1,1,1,1,1,1,1,y),~\mathcal{T}_4=\diag(y^{-1/2},y^{-1/2},1,y^{1/2},y^{1/2})$.
In any case, as long as we do not violate ANZ strips, we can dispose of these
shifts conveniently. As for the kernel matrix, it can be divided into sixteen
submatrices as
\begin{align}
\mathbf{K}^{(5)}(x)=\begin{bmatrix}
\mathbf{K}^{(5)}_{1,1}(x) & \mathbf{K}^{(5)}_{1,2}(x) &
\mathbf{K}^{(5)}_{1,3}(x) &
\mathbf{K}^{(5)}_{1,4}(k) \\
\mathbf{K}^{(5)}_{2,1}(x) & \mathbf{K}^{(5)}_{2,2}(x) &
\mathbf{K}^{(5)}_{2,3}(x) &
\mathbf{K}^{(5)}_{2,4}(x) \\
\mathbf{K}^{(5)}_{3,1}(x) & \mathbf{K}^{(5)}_{3,2}(x) &
\mathbf{K}^{(5)}_{3,3}(x) &
\mathbf{K}^{(5)}_{3,4}(x)\\
\mathbf{K}^{(5)}_{4,1}(x) & \mathbf{K}^{(5)}_{4,2}(x) &
\mathbf{K}^{(5)}_{4,3}(x) &
\mathbf{K}^{(5)}_{4,4}(x)
\end{bmatrix}.\label{ker-su5}
\end{align}
\begin{align}
	\mathbf{K}^{(5)}(x) = \left[\mathbf{K}^{(5)}(x)\right]^{\dagger}, &&
\mathbf{K}^{(5)}_{i,j}(x)= {\cal T}_i^{-1} \Pi_{i} {\cal T}_{5-i}^{-1} {\left[\mathbf{K}^{(5) }_{5-j,5-i}(x)\right]}^t {\cal T}_{5-j} \Pi_{j} {\cal T}_{j},
\end{align}
where we have defined the reflection matrix $\left[\Pi_{j}\right]_{mn}= \delta_{d^{(5)}_j+1-m,n} $.
This way, from the sixteen submatrices we must only treat
six of them explicitly. Indeed, in Fourier space they are given as
\begin{align}
\hat{\mathbf{K}}^{(5)}_{1,1}(k)&={\cal T}_1^{-1}\begin{bmatrix}
\hat{K}^{(5)}_0 & \hat{K}^{(5)}_1 & \hat{K}^{(5)}_1 & \hat{K}^{(5)}_1 & \hat{K}^{(5)}_1 \\
\hat{K}^{(5)}_2 & \hat{K}^{(5)}_0 & \hat{K}^{(5)}_1 & \hat{K}^{(5)}_1 & \hat{K}^{(5)}_1 \\
\hat{K}^{(5)}_2 & \hat{K}^{(5)}_2 & \hat{K}^{(5)}_0 & \hat{K}^{(5)}_1 & \hat{K}^{(5)}_1 \\
\hat{K}^{(5)}_2 & \hat{K}^{(5)}_2 & \hat{K}^{(5)}_2 & \hat{K}^{(5)}_0 & \hat{K}^{(5)}_1 \\
\hat{K}^{(5)}_2 & \hat{K}^{(5)}_2 & \hat{K}^{(5)}_2 & \hat{K}^{(5)}_2 & \hat{K}^{(5)}_0
\end{bmatrix}{\cal T}_1,
\end{align}
where
\begin{align}
\hat{K}^{(5)}_0(k)&=\mathcal{K}_{[5]}^{(1,1)}(k), \\
\hat{K}^{(5)}_1(k)&=\mathcal{K}_{[5]}^{(1,1)}(k)+\ee^{-k/2-|k|/2}, \\
\hat{K}^{(5)}_2(k)&=\mathcal{K}_{[5]}^{(1,1)}(k)+\ee^{k/2-|k|/2}.
\end{align}
\begin{align}
\hat{\mathbf{K}}^{(5)}_{1,2}(k)&={\cal T}_1^{-1}\begin{bmatrix}[1.6]
\hat{K}^{(5)}_3 & \hat{K}^{(5)}_3 & \hat{K}^{(5)}_6 & \hat{K}^{(5)}_3 & \hat{K}^{(5)}_3 & \hat{K}^{(5)}_6 & \hat{K}^{(5)}_6 & \hat{K}^{(5)}_6 & \hat{K}^{(5)}_6 & \hat{K}^{(5)}_6 \\
\hat{K}^{(5)}_3 & \hat{K}^{(5)}_5 & \hat{K}^{(5)}_3 & \hat{K}^{(5)}_5 & \hat{K}^{(5)}_5 & \hat{K}^{(5)}_3 & \hat{K}^{(5)}_3 & \hat{K}^{(5)}_6 & \hat{K}^{(5)}_6 & \hat{K}^{(5)}_6 \\
\hat{K}^{(5)}_4 & \hat{K}^{(5)}_3 & \hat{K}^{(5)}_3 & \hat{K}^{(5)}_5 & \hat{K}^{(5)}_5 & \hat{K}^{(5)}_5 & \hat{K}^{(5)}_5 & \hat{K}^{(5)}_3 & \hat{K}^{(5)}_3 & \hat{K}^{(5)}_6 \\
\hat{K}^{(5)}_4 & \hat{K}^{(5)}_4 & \hat{K}^{(5)}_4 & \hat{K}^{(5)}_3 & \hat{K}^{(5)}_5 & \hat{K}^{(5)}_3 & \hat{K}^{(5)}_5 & \hat{K}^{(5)}_3 & \hat{K}^{(5)}_5 & \hat{K}^{(5)}_3 \\
\hat{K}^{(5)}_4 & \hat{K}^{(5)}_4 & \hat{K}^{(5)}_4 & \hat{K}^{(5)}_4 & \hat{K}^{(5)}_3 & \hat{K}^{(5)}_4 & \hat{K}^{(5)}_3 & \hat{K}^{(5)}_4 & \hat{K}^{(5)}_3 & \hat{K}^{(5)}_3
\end{bmatrix}{\cal T}_2,
\end{align}
where
\begin{align}
\hat{K}^{(5)}_3(k)&=\mathcal{K}_{[5]}^{(1,2)}(k), \\
\hat{K}^{(5)}_4(k)&=\mathcal{K}_{[5]}^{(1,2)}(k)+\ee^{k-|k|/2}-\ee^{k/2}, \\
\hat{K}^{(5)}_5(k)&=\mathcal{K}_{[5]}^{(1,2)}(k)+\ee^{-|k|/2}, \\
\hat{K}^{(5)}_6(k)&=\mathcal{K}_{[5]}^{(1,2)}(k)+\ee^{-k-|k|/2}-\ee^{-k/2},
\end{align}
\begin{align}
\mathbf{\hat{K}}_{1,3}^{(5)}(k)&={\cal T}_1^{-1}\begin{bmatrix}[1.6]
\hat{K}^{(5)}_7 & \hat{K}^{(5)}_7 & \hat{K}^{(5)}_7 & \hat{K}^{(5)}_7 & \hat{K}^{(5)}_7 & \hat{K}^{(5)}_{11} & \hat{K}^{(5)}_{11} & \hat{K}^{(5)}_7 & \hat{K}^{(5)}_{11} & \hat{K}^{(5)}_{11} \\
\hat{K}^{(5)}_7 & \hat{K}^{(5)}_7 & \hat{K}^{(5)}_7 & \hat{K}^{(5)}_{10} & \hat{K}^{(5)}_{10} & \hat{K}^{(5)}_7 & \hat{K}^{(5)}_7 & \hat{K}^{(5)}_{10} & \hat{K}^{(5)}_7 & \hat{K}^{(5)}_{11} \\
\hat{K}^{(5)}_7 & \hat{K}^{(5)}_9 & \hat{K}^{(5)}_9 & \hat{K}^{(5)}_7 & \hat{K}^{(5)}_7 & \hat{K}^{(5)}_7 & \hat{K}^{(5)}_7 & \hat{K}^{(5)}_{10} & \hat{K}^{(5)}_{10} & \hat{K}^{(5)}_7 \\
\hat{K}^{(5)}_8 & \hat{K}^{(5)}_7 & \hat{K}^{(5)}_9 & \hat{K}^{(5)}_7 & \hat{K}^{(5)}_9 & \hat{K}^{(5)}_7 & \hat{K}^{(5)}_9 & \hat{K}^{(5)}_7 & \hat{K}^{(5)}_7 & \hat{K}^{(5)}_7 \\
\hat{K}^{(5)}_8 & \hat{K}^{(5)}_8 & \hat{K}^{(5)}_7 & \hat{K}^{(5)}_8 & \hat{K}^{(5)}_7 & \hat{K}^{(5)}_8 & \hat{K}^{(5)}_7 & \hat{K}^{(5)}_7 & \hat{K}^{(5)}_7 & \hat{K}^{(5)}_7
\end{bmatrix}{\cal T}_3,
\end{align}
where
\begin{align}
\hat{K}^{(5)}_{7}(k)&=\mathcal{K}_{[5]}^{(1,3)}(k), \\
\hat{K}^{(5)}_{8}(k)&=\mathcal{K}_{[5]}^{(1,3)}(k)+\ee^{3k/2-|k|/2}-\ee^{k}, \\
\hat{K}^{(5)}_{9}(k)&=\mathcal{K}_{[5]}^{(1,3)}(k)+\ee^{k/2-|k|/2}, \\
\hat{K}^{(5)}_{10}(k)&=\mathcal{K}_{[5]}^{(1,3)}(k)+\ee^{-k/2-|k|/2}, \\
\hat{K}^{(5)}_{11}(k)&=\mathcal{K}_{[5]}^{(1,3)}(k)+\ee^{-3k/2-|k|/2}-\ee^{-k},
\end{align}
\begin{align}
\mathbf{\hat{K}}_{1,4}^{(5)}(k)&={\cal T}_1^{-1}\begin{bmatrix}[1.6]
\hat{K}^{(5)}_{12} & \hat{K}^{(5)}_{12} & \hat{K}^{(5)}_{12} & \hat{K}^{(5)}_{12} & \hat{K}^{(5)}_{17} \\
\hat{K}^{(5)}_{12} & \hat{K}^{(5)}_{12} & \hat{K}^{(5)}_{12} & \hat{K}^{(5)}_{16} & \hat{K}^{(5)}_{12} \\
\hat{K}^{(5)}_{12} & \hat{K}^{(5)}_{12} & \hat{K}^{(5)}_{15} & \hat{K}^{(5)}_{12} & \hat{K}^{(5)}_{12} \\
\hat{K}^{(5)}_{12} & \hat{K}^{(5)}_{14} & \hat{K}^{(5)}_{12} & \hat{K}^{(5)}_{12} & \hat{K}^{(5)}_{12} \\
\hat{K}^{(5)}_{13} & \hat{K}^{(5)}_{12} & \hat{K}^{(5)}_{12} & \hat{K}^{(5)}_{12} & \hat{K}^{(5)}_{12}
\end{bmatrix}{\cal T}_4,
\end{align}
where
\begin{align}
\hat{K}^{(5)}_{12}(k)&=\mathcal{K}_{[5]}^{(1,4)}(k), \\
\hat{K}^{(5)}_{13}(k)&=\mathcal{K}_{[5]}^{(1,4)}(k)+\ee^{2k-|k|/2}-\ee^{3k/2}, \\
\hat{K}^{(5)}_{14}(k)&=\mathcal{K}_{[5]}^{(1,4)}(k)+\ee^{k-|k|/2}, \\
\hat{K}^{(5)}_{15}(k)&=\mathcal{K}_{[5]}^{(1,4)}(k)+\ee^{-|k|/2}, \\
\hat{K}^{(5)}_{16}(k)&=\mathcal{K}_{[5]}^{(1,4)}(k)+\ee^{-k-|k|/2}, \\
\hat{K}^{(5)}_{17}(k)&=\mathcal{K}_{[5]}^{(1,4)}(k)+\ee^{-2k-|k|/2}-\ee^{-3k/2},
\end{align}
\begin{align}
\mathbf{\hat{K}}_{2,2}^{(5)}(k)&={\cal T}_2^{-1}\begin{bmatrix}[1.6]
\hat{K}^{(5)}_{18} & \hat{K}^{(5)}_{21} & \hat{K}^{(5)}_{21} & \hat{K}^{(5)}_{21} &
\hat{K}^{(5)}_{21} &
\hat{K}^{(5)}_{21} & \hat{K}^{(5)}_{21} & \hat{K}^{(5)}_{25} & \hat{K}^{(5)}_{25} & \hat{K}^{(5)}_{25} \\
\hat{K}^{(5)}_{19} & \hat{K}^{(5)}_{18} & \hat{K}^{(5)}_{21} & \hat{K}^{(5)}_{21} & \hat{K}^{(5)}_{21} & \hat{K}^{(5)}_{24} & \hat{K}^{(5)}_{24} & \hat{K}^{(5)}_{21} & \hat{K}^{(5)}_{21} & \hat{K}^{(5)}_{25} \\
\hat{K}^{(5)}_{19} & \hat{K}^{(5)}_{19} & \hat{K}^{(5)}_{18} & \hat{K}^{(5)}_{23} & \hat{K}^{(5)}_{23} & \hat{K}^{(5)}_{21} & \hat{K}^{(5)}_{21} & \hat{K}^{(5)}_{21} & \hat{K}^{(5)}_{21} & \hat{K}^{(5)}_{25} \\
\hat{K}^{(5)}_{19} & \hat{K}^{(5)}_{19} & \hat{K}^{(5)}_{23} & \hat{K}^{(5)}_{18} & \hat{K}^{(5)}_{21} & \hat{K}^{(5)}_{21} & \hat{K}^{(5)}_{24} & \hat{K}^{(5)}_{21} & \hat{K}^{(5)}_{24} & \hat{K}^{(5)}_{21} \\
\hat{K}^{(5)}_{19} & \hat{K}^{(5)}_{19} & \hat{K}^{(5)}_{23} & \hat{K}^{(5)}_{19} & \hat{K}^{(5)}_{18} & \hat{K}^{(5)}_{23} & \hat{K}^{(5)}_{21} & \hat{K}^{(5)}_{23} & \hat{K}^{(5)}_{21} & \hat{K}^{(5)}_{21} \\
\hat{K}^{(5)}_{19} & \hat{K}^{(5)}_{22} & \hat{K}^{(5)}_{19} & \hat{K}^{(5)}_{19} & \hat{K}^{(5)}_{23} & \hat{K}^{(5)}_{18} & \hat{K}^{(5)}_{21} & \hat{K}^{(5)}_{21} & \hat{K}^{(5)}_{24} & \hat{K}^{(5)}_{21} \\
\hat{K}^{(5)}_{19} & \hat{K}^{(5)}_{22} & \hat{K}^{(5)}_{19} & \hat{K}^{(5)}_{22} & \hat{K}^{(5)}_{19} & \hat{K}^{(5)}_{19} & \hat{K}^{(5)}_{18} & \hat{K}^{(5)}_{23} & \hat{K}^{(5)}_{21} & \hat{K}^{(5)}_{21} \\
\hat{K}^{(5)}_{20} & \hat{K}^{(5)}_{19} & \hat{K}^{(5)}_{19} & \hat{K}^{(5)}_{19} & \hat{K}^{(5)}_{23} & \hat{K}^{(5)}_{19} & \hat{K}^{(5)}_{23} & \hat{K}^{(5)}_{18} & \hat{K}^{(5)}_{21} & \hat{K}^{(5)}_{21} \\
\hat{K}^{(5)}_{20} & \hat{K}^{(5)}_{19} & \hat{K}^{(5)}_{19} & \hat{K}^{(5)}_{22} & \hat{K}^{(5)}_{19} & \hat{K}^{(5)}_{22} & \hat{K}^{(5)}_{19} & \hat{K}^{(5)}_{19} & \hat{K}^{(5)}_{18} & \hat{K}^{(5)}_{21} \\
\hat{K}^{(5)}_{20} & \hat{K}^{(5)}_{20} & \hat{K}^{(5)}_{20} & \hat{K}^{(5)}_{19} & \hat{K}^{(5)}_{19} & \hat{K}^{(5)}_{19} & \hat{K}^{(5)}_{19} & \hat{K}^{(5)}_{19} & \hat{K}^{(5)}_{19} & \hat{K}^{(5)}_{18}
\end{bmatrix}{\cal T}_2,
\end{align}
where
\begin{align}
\hat{K}^{(5)}_{18}(k)&=\mathcal{K}_{[5]}^{(2,2)}(k), \\
\hat{K}^{(5)}_{19}(k)&=\mathcal{K}_{[5]}^{(2,2)}(k)+\ee^{k/2-|k|/2}, \\
\hat{K}^{(5)}_{20}(k)&=\mathcal{K}_{[5]}^{(2,2)}(k)+\ee^{k-|k|}-\ee^{k}, \\
\hat{K}^{(5)}_{21}(k)&=\mathcal{K}_{[5]}^{(2,2)}(k)+\ee^{-k/2-|k|/2}, \\
\hat{K}^{(5)}_{22}(k)&=\mathcal{K}_{[5]}^{(2,2)}(k)+2\ee^{k/2-|k|/2}, \\
\hat{K}^{(5)}_{23}(k)&=\mathcal{K}_{[5]}^{(2,2)}(k)+\ee^{-|k|}, \\
\hat{K}^{(5)}_{24}(k)&=\mathcal{K}_{[5]}^{(2,2)}(k)+2\ee^{-k/2-|k|/2}, \\
\hat{K}^{(5)}_{25}(k)&=\mathcal{K}_{[5]}^{(2,2)}(k)+\ee^{-k-|k|}-\ee^{-k},
\end{align}
\begin{align}
\mathbf{\hat{K}}_{2,3}^{(5)}(k)&={\cal T}_2^{-1}\begin{bmatrix}[1.6]
\hat{K}^{(5)}_{26} & \hat{K}^{(5)}_{26} & \hat{K}^{(5)}_{26} & \hat{K}^{(5)}_{29} & \hat{K}^{(5)}_{29} & \hat{K}^{(5)}_{29} & \hat{K}^{(5)}_{29} & \hat{K}^{(5)}_{29} & \hat{K}^{(5)}_{29} & \hat{K}^{(5)}_{39} \\
\hat{K}^{(5)}_{26} & \hat{K}^{(5)}_{28} & \hat{K}^{(5)}_{28} & \hat{K}^{(5)}_{26} & \hat{K}^{(5)}_{26} & \hat{K}^{(5)}_{29} & \hat{K}^{(5)}_{29} & \hat{K}^{(5)}_{29} & \hat{K}^{(5)}_{38} & \hat{K}^{(5)}_{29} \\
\hat{K}^{(5)}_{26} & \hat{K}^{(5)}_{28} & \hat{K}^{(5)}_{28} & \hat{K}^{(5)}_{28} & \hat{K}^{(5)}_{28} & \hat{K}^{(5)}_{26} & \hat{K}^{(5)}_{26} & \hat{K}^{(5)}_{37} & \hat{K}^{(5)}_{29} & \hat{K}^{(5)}_{29} \\
\hat{K}^{(5)}_{27} & \hat{K}^{(5)}_{26} & \hat{K}^{(5)}_{28} & \hat{K}^{(5)}_{26} & \hat{K}^{(5)}_{28} & \hat{K}^{(5)}_{29} & \hat{K}^{(5)}_{36} & \hat{K}^{(5)}_{26} & \hat{K}^{(5)}_{29} & \hat{K}^{(5)}_{29} \\
\hat{K}^{(5)}_{27} & \hat{K}^{(5)}_{27} & \hat{K}^{(5)}_{26} & \hat{K}^{(5)}_{27} & \hat{K}^{(5)}_{26} & \hat{K}^{(5)}_{35} & \hat{K}^{(5)}_{29} & \hat{K}^{(5)}_{26} & \hat{K}^{(5)}_{29} & \hat{K}^{(5)}_{29} \\
\hat{K}^{(5)}_{27} & \hat{K}^{(5)}_{26} & \hat{K}^{(5)}_{28} & \hat{K}^{(5)}_{28} & \hat{K}^{(5)}_{34} & \hat{K}^{(5)}_{26} & \hat{K}^{(5)}_{28} & \hat{K}^{(5)}_{28} & \hat{K}^{(5)}_{26} & \hat{K}^{(5)}_{29} \\
\hat{K}^{(5)}_{27} & \hat{K}^{(5)}_{27} & \hat{K}^{(5)}_{26} & \hat{K}^{(5)}_{33} & \hat{K}^{(5)}_{28} & \hat{K}^{(5)}_{27} & \hat{K}^{(5)}_{26} & \hat{K}^{(5)}_{28} & \hat{K}^{(5)}_{26} & \hat{K}^{(5)}_{29} \\
\hat{K}^{(5)}_{27} & \hat{K}^{(5)}_{27} & \hat{K}^{(5)}_{32} & \hat{K}^{(5)}_{26} & \hat{K}^{(5)}_{28} & \hat{K}^{(5)}_{26} & \hat{K}^{(5)}_{28} & \hat{K}^{(5)}_{28} & \hat{K}^{(5)}_{28} & \hat{K}^{(5)}_{26} \\
\hat{K}^{(5)}_{27} & \hat{K}^{(5)}_{31} & \hat{K}^{(5)}_{27} & \hat{K}^{(5)}_{27} & \hat{K}^{(5)}_{26} & \hat{K}^{(5)}_{27} & \hat{K}^{(5)}_{26} & \hat{K}^{(5)}_{28} & \hat{K}^{(5)}_{28} & \hat{K}^{(5)}_{26} \\
\hat{K}^{(5)}_{30} & \hat{K}^{(5)}_{27} & \hat{K}^{(5)}_{27} & \hat{K}^{(5)}_{27} & \hat{K}^{(5)}_{27} & \hat{K}^{(5)}_{27} & \hat{K}^{(5)}_{27} & \hat{K}^{(5)}_{26} & \hat{K}^{(5)}_{26} & \hat{K}^{(5)}_{26}
\end{bmatrix}{\cal T}_3,
\end{align}
where
\begin{align}
\hat{K}^{(5)}_{26}(k)&=\mathcal{K}_{[5]}^{(2,3)}(k), \\
\hat{K}^{(5)}_{27}(k)&=\mathcal{K}_{[5]}^{(2,3)}(k)+\ee^{k-|k|/2}-\ee^{k/2}, \\
\hat{K}^{(5)}_{28}(k)&=\mathcal{K}_{[5]}^{(2,3)}(k)+\ee^{-|k|/2}, \\
\hat{K}^{(5)}_{29}(k)&=\mathcal{K}_{[5]}^{(2,3)}(k)+\ee^{-k-|k|/2}-\ee^{-k/2}, \\
\hat{K}^{(5)}_{30}(k)&=\mathcal{K}_{[5]}^{(2,3)}(k)+\ee^{3k/2-|k|}-\ee^{3k/2}-\ee^{k/2}, \\
\hat{K}^{(5)}_{31}(k)&=\mathcal{K}_{[5]}^{(2,3)}(k)+2\ee^{k-|k|/2}-\ee^{k/2}, \\
\hat{K}^{(5)}_{32}(k)&=\mathcal{K}_{[5]}^{(2,3)}(k)+\ee^{k/2-|k|}-\ee^{k/2}, \\
\hat{K}^{(5)}_{33}(k)&=\mathcal{K}_{[5]}^{(2,3)}(k)+\ee^{k/2-|k|}, \\
\hat{K}^{(5)}_{34}(k)&=\mathcal{K}_{[5]}^{(2,3)}(k)+2\ee^{-|k|/2}, \\
\hat{K}^{(5)}_{35}(k)&=\mathcal{K}_{[5]}^{(2,3)}(k)+\ee^{-3|k|/2}-\ee^{-|k|/2}, \\
\hat{K}^{(5)}_{36}(k)&=\mathcal{K}_{[5]}^{(2,3)}(k)+\ee^{-k/2-|k|}, \\
\hat{K}^{(5)}_{37}(k)&=\mathcal{K}_{[5]}^{(2,3)}(k)+\ee^{-k/2-|k|}-\ee^{-k/2}, \\
\hat{K}^{(5)}_{38}(k)&=\mathcal{K}_{[5]}^{(2,3)}(k)+2\ee^{-k-|k|/2}-\ee^{-k/2}, \\
\hat{K}^{(5)}_{39}(k)&=\mathcal{K}_{[5]}^{(2,3)}(k)+\ee^{-3k/2-|k|}-\ee^{-3k/2}-\ee^{-k/2},
\end{align}
where we used again the common function defined on Eq.\eqref{common} with $n=5$. Lastly, the largest eigenvalue of the QTM is given by
\eq
\lim_{N \rightarrow \infty}\log \frac{\Lambda_{1,1}^{(5)}(x)}{\Phi_+(x-\im) \Phi_-(x+\im)} = -\im \beta J\partial_x \log \frac{\Gamma(
    \frac{1}{5}- \im \frac{x}{5}) \Gamma(1 + \im \frac{x}{5})}{\Gamma(\frac{1}{5} + \im \frac{x}{5}) \Gamma(1 - \im \frac{x}{5})} + \beta \sum_{j=1}^{5}\frac{\mu_j}{5} + \mathbf{d}^{(5)}(x)^{\dagger} \ast \log \mathbf{B}^{(5)}(x).
\en
from which one obtains the thermodynamic properties at $x=0$.

\section{Models and Numerics}\label{NUMERICS}
\begin{figure}[h!]
  \centering
   \includegraphics[width=6.5cm]{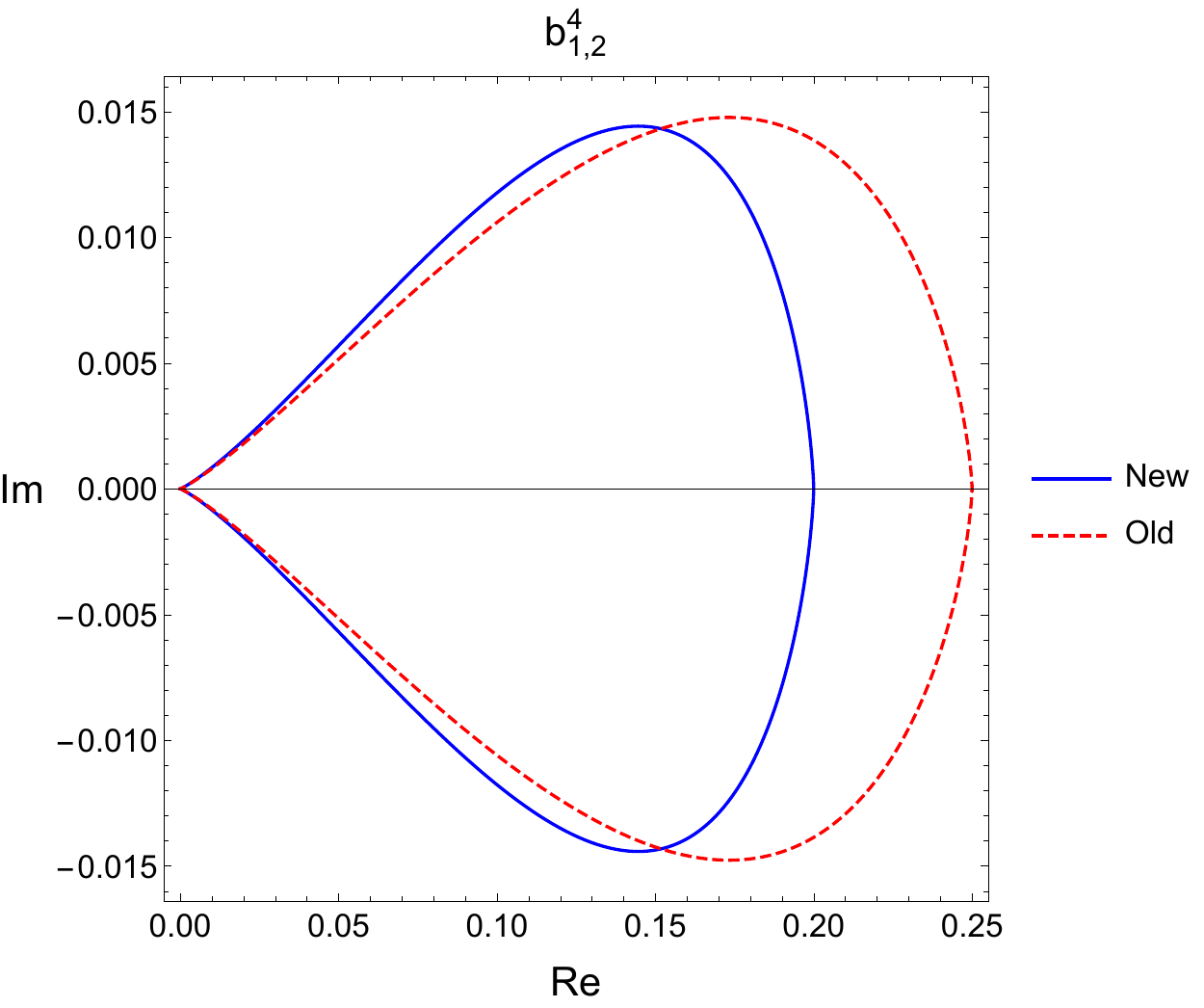}
   \includegraphics[width=6.5cm]{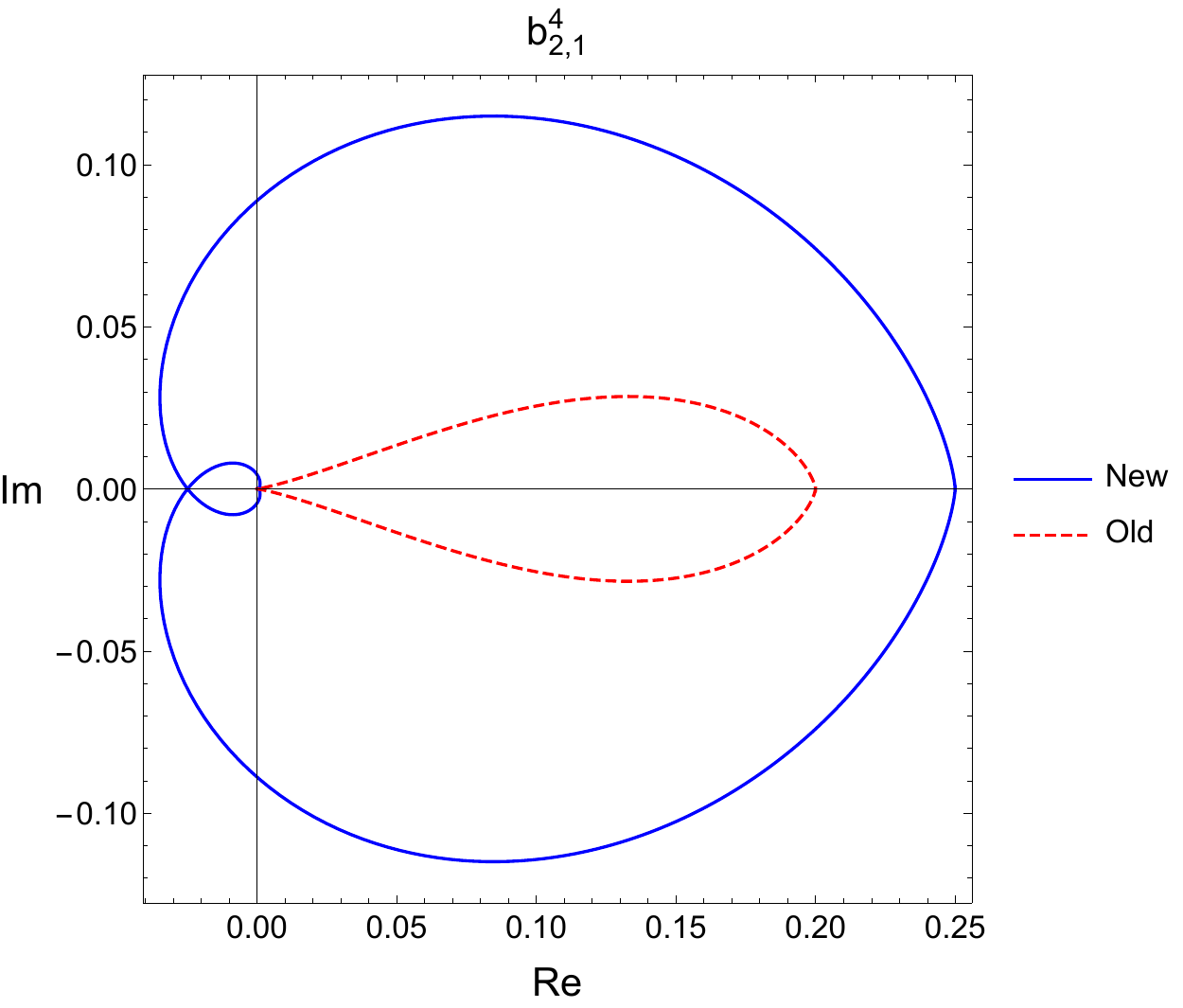}
   \includegraphics[width=6.5cm]{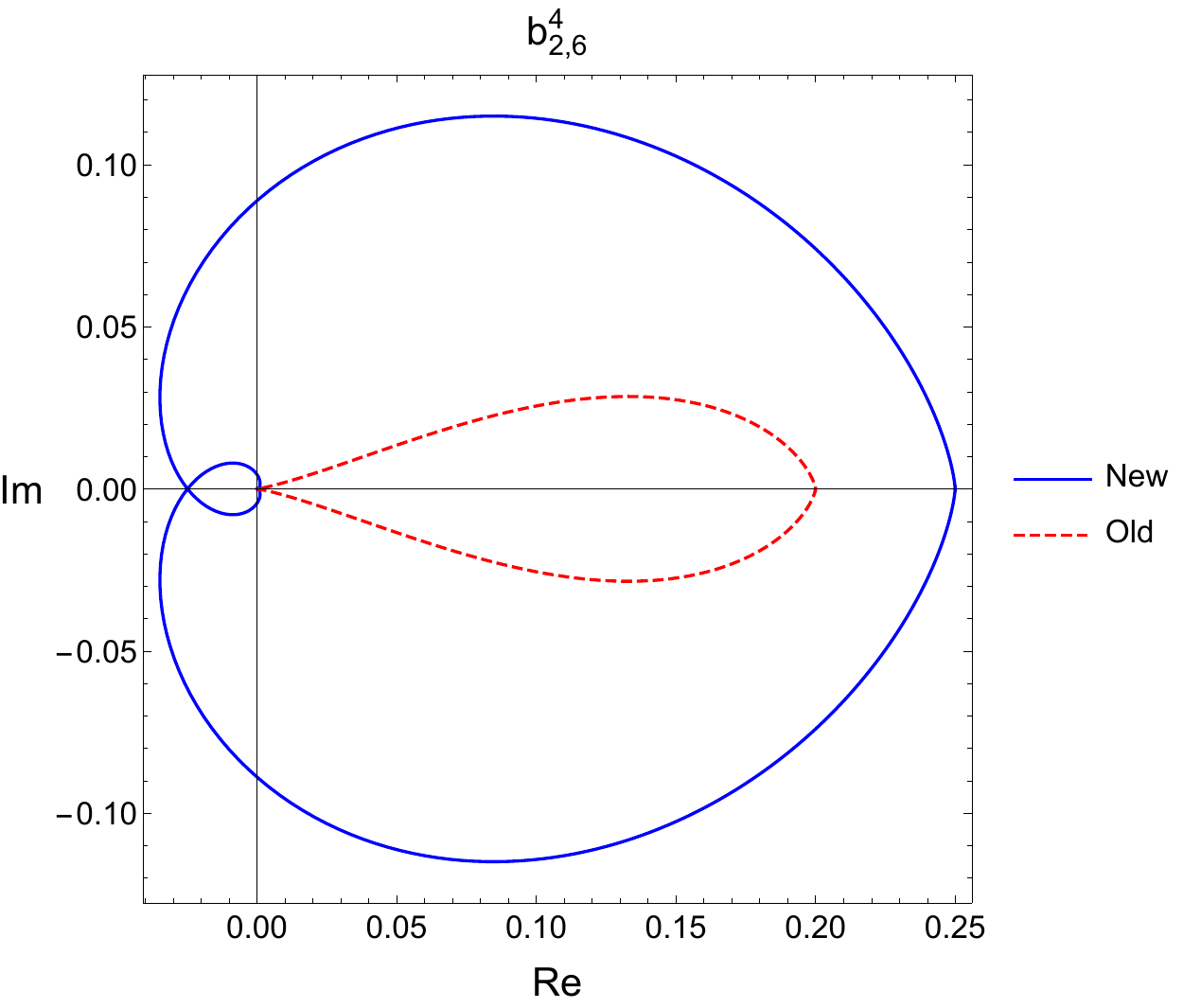}
   \includegraphics[width=6.5cm]{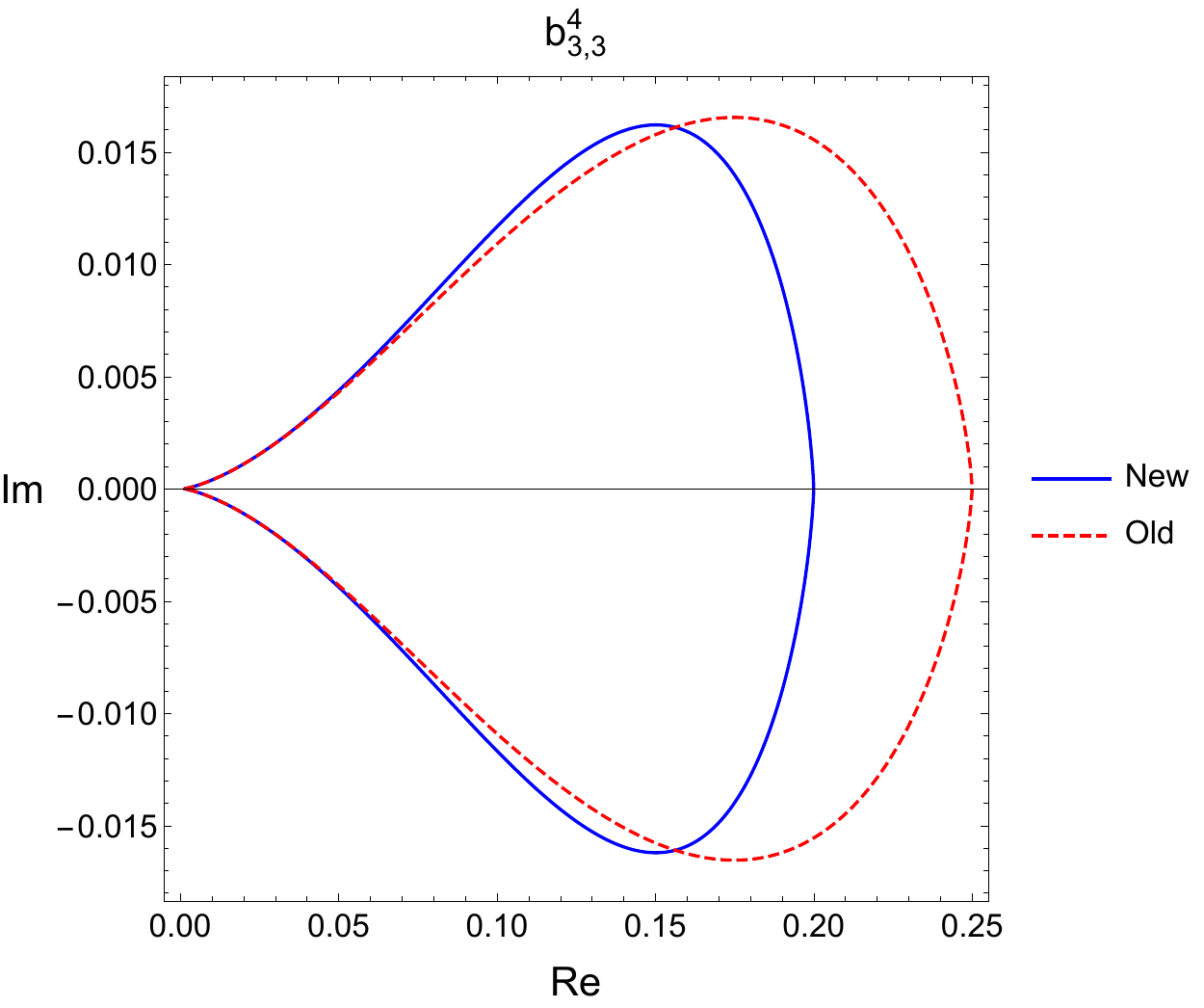}
  \caption{
  Traces of the curves of the canonical auxiliary functions that differ from the set of functions in \cite{DAMERAU}. $T=0.1,~\mu=0$.
    }\label{sl4Canonical} \label{curvesCanonical}
\end{figure}

For lower rank models,  $sl_{2}$ and $sl_{3}$, there are no differences between our
proposed set of auxiliary functions and the successful approaches in the
past\cite{KLUMPER-SU3}. Nevertheless, with respect to the $sl_4$ we can already see that the canonical functions here proposed by us differ from \cite{DAMERAU}.
Both formulations allow for stable and fast algorithms, so in terms of numerical efficiency there is little difference. Using the same set of physical parameters (temperature and
chemical potentials), as well as for the iterative solution of the NLIEs
(discretization parameters in Fast Fourier Transform), we obtained the solution for the canonical functions
after 42 iterations, whereas solutions for the old formulation were obtained after 38 iterations, Fig.\ref{curvesCanonical}.
We see that
  the functions in question have the following advantages and disadvantages.
  The definition of the canonical set of functions is obtained in a systematic
  manner, however some functions take a slightly more involved trace
  accounting for slightly longer computation times in comparison to the
  functions used in \cite{KLUMPER-SU3,DAMERAU} for which there is no systematic
  construction principle as to now.

Nevertheless, this tradeoff reveals to be worthwhile, for now we may tackle
the $sl_5$ case. Here we show that our approach works by displaying physical
properties for some particular choices of physical parameters. When all
chemical potentials are equal we have a balanced distribution
$n_i=\frac{1}{5},\ i=1,\ldots,5$. The physical significance of such an
assertion depends on how we interpret the model. If we regard the permutation
operator as the Hamiltonian describing exchange interaction in
a spin chain with $ 2 S+1=5$,
\begin{equation}
{\cal H}= \sum_{i=1}^L -\frac{5}{2} \vec{S}_i \cdot \vec{S}_{i+1}-\frac{13}{36} {\left(\vec{S}_i \cdot \vec{S}_{i+1}\right)}^2+\frac{1}{6} {\left(\vec{S}_i \cdot \vec{S}_{i+1}\right)}^3+\frac{1}{36} {\left(\vec{S}_i \cdot \vec{S}_{i+1}\right)}^4 ,
\end{equation}
then equal species densities  result especially in zero magnetization. Nevertheless,
here we propose the following Hamiltonian
\begin{equation}
{\cal H}= \sum_{i=1}^L {\cal P}\left[ \sum_{\alpha=1}^4 -\left(b_{\alpha,i}^\dagger b_{\alpha,i+1}+b_{\alpha,i+1}^\dagger b_{\alpha,i}\right) +2 n_{\alpha,i} n_{\alpha,i+1}+ \sum_{\alpha \neq \beta=1}^4 n_{\alpha,i} n_{\beta, i+1}+ b_{\alpha,i+1}^\dagger b_{\beta,i}^\dagger b_{\beta,i+1} b_{\alpha,i}\right] {\cal P}, \label{hardcore}
\end{equation}
which describes hard core bosons of four types with
  nearest-neighbour hopping and interactions on the chain. The operators
  $b_{\alpha,i}$ are bosonic in nature and may correspond to pairs of
  electrons that can be in triplet or singlet states, corresponding to the
  four possibilities of the indices $\alpha,~\beta$. Furthermore, one of the
  five local states plays the role of the vacant site. In this case, equal species
density means that we have a fraction of holes of $\frac{1}{5}$ which makes
the density of the pairs to be $\frac{4}{5}$ with no net magnetization.

\begin{figure}[h!]
  \centering
   \includegraphics[width=7cm]{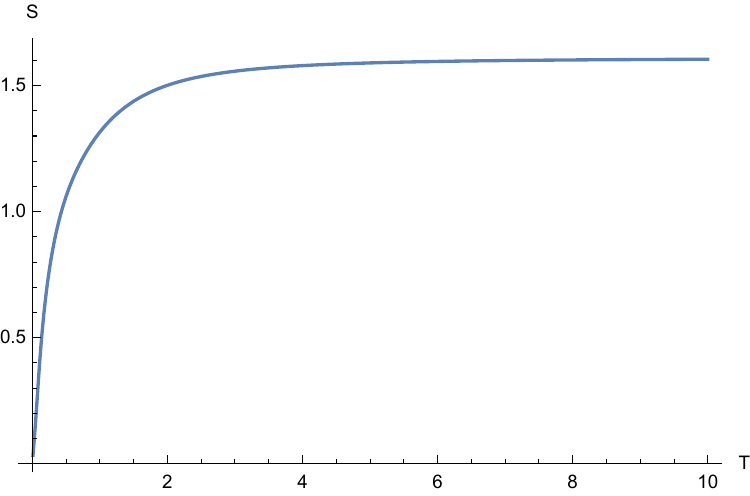}
   \includegraphics[width=7cm]{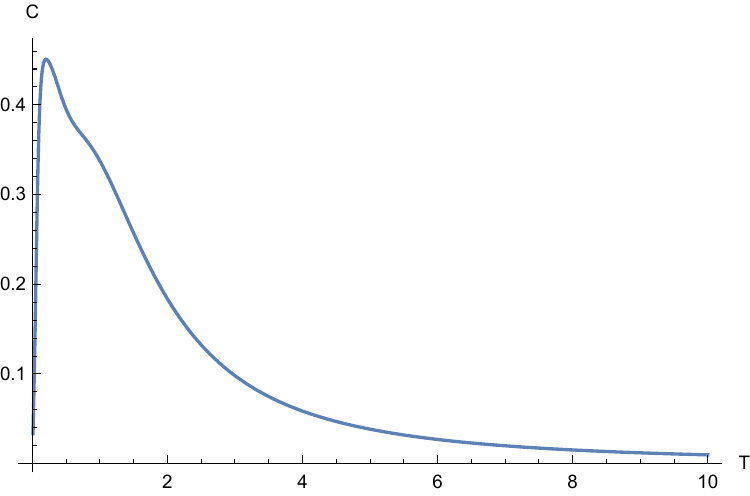}
  \caption{ Entropy and specific-heat at pair density  $\frac{4}{5}$}\label{FigsT}
\end{figure}

From the numerics, we can see that the system reaches the expected high and
low temperature limits with respect to the entropy and specific heat, see
Fig.\ref{FigsT}. For high temperature all states become equally probable and
we have $S = \log 5$, whereas for the specific heat we find the Schottky
anomaly, i.e.~the specific heat approaches zero for low and high
  temperatures and thus has at least one temperature maximum. This is typical
 of lattice systems with a finite degree of freedom per site.

Although we keep the chemical potentials equal, we can evaluate derivatives of
the thermodynamical potential with respect to them, leading to the
compressibility
$\chi_{i}=\left(\frac{\partial n_i}{\partial
	\mu_i}\right)_{T, \mu \backslash i}$ and convertibility
$\chi_{i,j}=-\left(\frac{\partial n_j}{\partial \mu_i}\right)_{T, \mu \backslash i}$, where in the latter $j \neq i$ and ${T, \mu \backslash i}$
 is the set of all chemical potentials except for
  $\mu_i$. Because of symmetry, we have $\chi_i= 4
\chi_{i,j}$. In this sense, only one independent susceptibility exists from which the
physical properties such as usual susceptibilities of the actual spin chain
may be obtained. Moreover, at equal
chemical potentials, we observe indeed that $\left(\frac{\partial
  n_i}{\partial T}\right)_{\boldmath \mu}=0$.
\begin{figure}[h!]
  \centering
   \includegraphics[width=7cm]{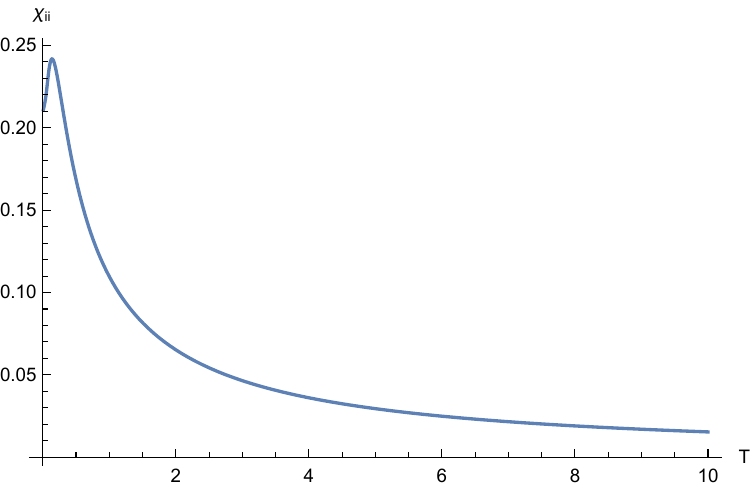}
   \includegraphics[width=7cm]{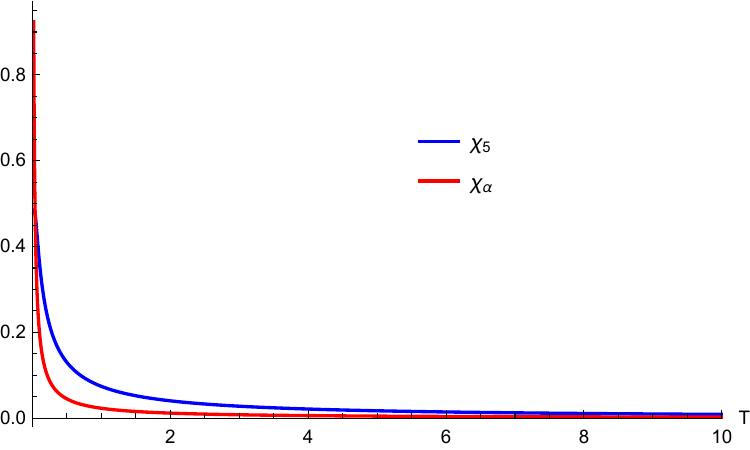}
   \caption{Compressibility at pair density  $\frac{4}{5}$ (a). Generalized susceptibilities at pair density  $0.1(b).$
   }\label{FigsMu}
\end{figure}

In contrast, as we move to lower pair densities, there arises an asymmetry
between hole and particle states. In view of this, two independent generalized
susceptibilities exist, which we take to be $\chi_5$ and
$\chi_{\alpha}$. For concreteness, we let the local state No. $5$ represent a vacancy, whereas $\alpha=1,~2,~3,~4$ represent local states populated by hard-core bosons in the corresponding particular configuration. Then $\chi_5$ provides the charge susceptibility, while $\chi_{\alpha}$ gives specific susceptibility of particle species $\alpha$.

Such a difference indicates a separation phenomenon between the charge degree of
freedom and the remaining ones, like the spin and possibly others. This is
also apparent when comparing specific heat profiles at high and low particle
densities.  One can observe that different types of elementary excitations
take part in this model. Specially at pair density 0.1, the emergence of a
second maximum seems to be imminent. Such structures were observed in the
context of the t-J model, where Luttinger liquid properties were
probed\cite{KLUMPER-TJ2}. In the present case, although our model is bosonic
in character and it explicitly ignores pair breaking processes it still shows
separation features.
\begin{figure}[h!]
  \centering
   \includegraphics[width=7cm]{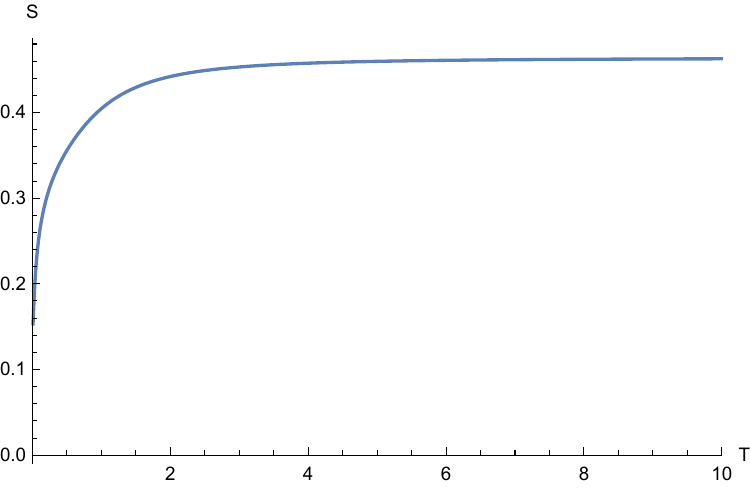}
   \includegraphics[width=7cm]{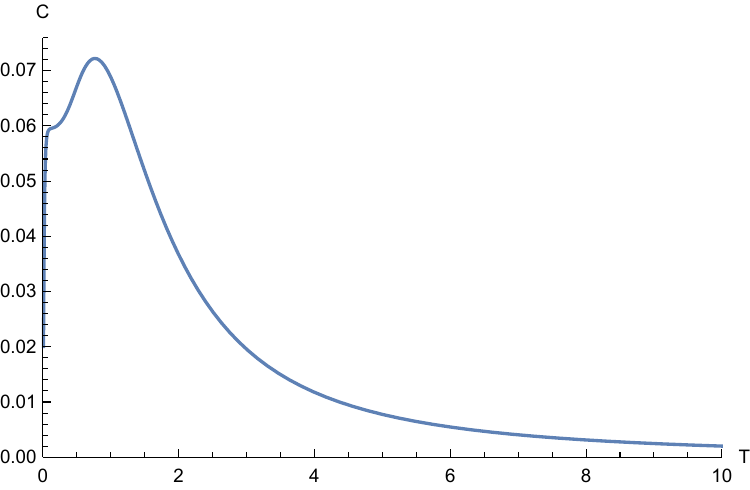}
  \caption{ Entropy and specific-heat at pair density  $0.1$}\label{FigsT2}
\end{figure}

\section{Closing remarks}
In our present understanding of the QTM approach to
  thermodynamics, NLIE have to be derived in order to work with the unusual
  discrete Bethe root patterns, namely discrete distributions with
  accumulation points, especially when considering the infinite Trotter
  limit. However, we bypass these problems by use of theorems for
functions of complex variables applied to certain auxiliary functions with
suitable analyticity properties. The actual
set up of such auxiliary functions is a problem by itself and, so far,
non-systematic procedures were used to define them. In this paper we have
presented a systematic procedure.

The problem can be separated into different stages, for not only
the auxiliary functions must have good analytical properties but
  also they must combine into sets that allow one to solve the resulting
  linear equations in Fourier space. Therefore the job is threefold:
  i) definition of auxiliary functions $\{B_j=1+b_j\}$
with same number of
  {\unknown}s that appear in their factorization, ii) analysis of their analytical
  properties, iii) expressing the {\unknown}s in terms of the uppercase auxiliary
  functions, and in turn expressing the lowercase auxiliary functions in terms
  of the {\unknown}s.

In this work we set out to generate functions i) systematically for
$sl_n$-invariant models as well as to derive the associated NLIEs. Our results
are satisfactory in applications at least up to $n=5$ which we
demonstrated by explicit solutions to the NLIE.  Our new route realizes a
previous conjecture according to which the number of auxiliary functions
should be $2^n-2$, while it abandons the heuristic guiding principle of
factorization of the Y-system \cite{DAMERAUT}.  Nevertheless, what is really
essential is the factorization of Young tableaux with higher symmetric spin
index. As a consequence, all {\unknown}s are column-like tableaux and they
come in finite number. There are differences between the previous approach and
our actual proposition for NLIE in case of the $sl_4$-invariant model, however
the new approach in detail recovers the previously derived NLIE for the
$sl_{2}$ and $sl_{3}$-invariant cases \cite{KLUMPER-SU3}.

We also elaborated on the determination of the {\unknown}s. Using
  pole cancellation graphs, we extracted the fundamental column tableaux that
  do not factorize. This further stimulated us to introduce a compact notation
  for the {\unknown}s. Naturally, many questions arise in this context. How many
auxiliary functions are there that can be written in terms of column tableaux?
Since there are many sets of equations for the same model, which one is the
most suitable? Do they all lead to a complete set of NLIEs? Can we still find
factorization for the Y-system? This addresses the question how
  to conform results with previous works. Factorization of the Y-system
  establishes the connection between the TBA and the QTM approach
  \cite{TAKAKLU}, and it would be desirable to carry out the program to
  higher-rank models.

  Besides the lack of factorization  of the Y-functions for
  $n>3$, another difficulty of the new method in the elementary form presented
  here concerns the analyticity of the auxiliary functions. Some of these
  functions contain {\unknown}s for which the distribution of zeros and poles
  requires a shift of the integration contour into the complex plane. It is
  not clear from the algebraic construction which ones or even how many
  functions require such a shift. In addition, these modifications introduce
  multiplicative factors that break some desired symmetries in the kernel
  matrix. We expect that working with functions based on Young tableaux with
  higher symmetric index $s$ solves the ``analyticity issue''.

To sum up, we have presented a machinery to generate what we call ``a
canonical set of functions''. However, as the questions posed by us indicate
these are far from being unique or even completely understood.

\section*{Acknowledgments}
All authors were supported by DFG through the research unit FOR 2316. I.R.P. and T.S.T. acknowledge financial support from Deutsche
Forschungsgemeinschaft through DFG projects KL 645/16-2 and Go 825/10-1. T.S.T. thanks G.A.P. Ribeiro for discussions.

\bibliographystyle{unsrt}
\bibliography{NLIEsl5-03-11-23}

\appendix
\section{Pole canceling graphs}\label{app-graphs}
As mentioned in Sec.\ref{GRAPHS}, the graphs that provide us with the adjacency matrices can also be used themselves to identify the {\unknown}s properly. We collect them by making perpendicular cuts on the edges of the graphs of each representation. The removed poles are the ones on the edges of the subgraphs while the nonremoved are the poles of the cut edges. Please notice one should not consider subgraphs that have already been collected from lower representations. Below we treat the $sl_4$ case.

By making all possible perpendicular cuts on Fig.\ref{sl4-1st}, we obtain for the first fundamental representation
\ytableausetup{boxsize=1.2em}
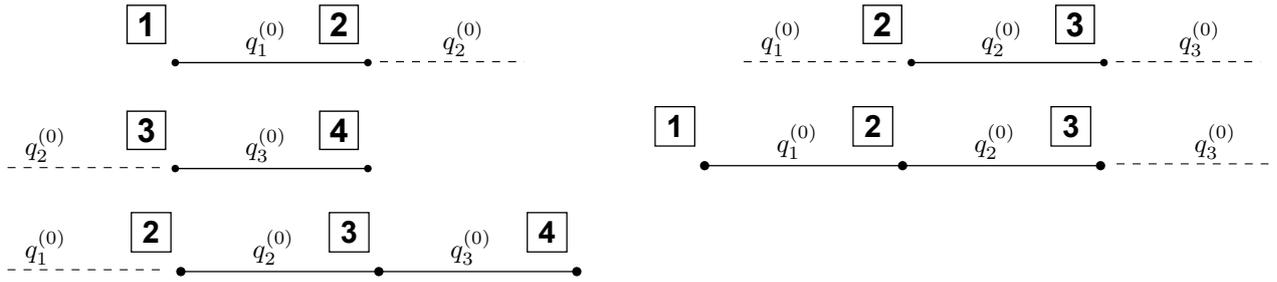
\begin{figure}[h!]
	\subfigure{\begin{tikzpicture}
		\node[anchor=south west,inner sep=0] at (0,-0.1) {\includegraphics[width=.2\textwidth]{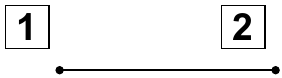}};
		\node at (1.9,0.3) {$q_1^{(0)}$};
		\draw[dashed] (3.4,0) -- (5.3,0);
		\node at (4.5,0.3) {$q_2^{(0)}$};
		\draw[dashed,white] (-1.5,0) -- (0.6,0);
		\end{tikzpicture}}
	\hfill
	\subfigure{\begin{tikzpicture}
		\node[anchor=south west,inner sep=0] at (0,-0.1) {\includegraphics[width=.2\textwidth]{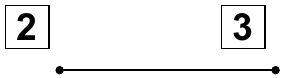}};
		\node at (1.9,0.3) {$q_2^{(0)}$};
		\draw[dashed] (3.4,0) -- (5.3,0);
		\node at (4.5,0.3) {$q_3^{(0)}$};
		\draw[dashed] (-1.5,0) -- (0.6,0);
		\node at (-1,0.3) {$q_1^{(0)}$};
		\end{tikzpicture}}
	\subfigure{\begin{tikzpicture}
		\node[anchor=south west,inner sep=0] at (0,-0.1) {\includegraphics[width=.2\textwidth]{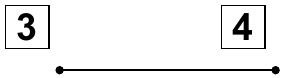}};
		\node at (1.9,0.3) {$q_3^{(0)}$};
		\draw[dashed] (-1.5,0) -- (0.6,0);
		\node at (-1,0.3) {$q_2^{(0)}$};
		\draw[dashed,white] (3.4,0) -- (5.3,0);
		\end{tikzpicture}}
	\subfigure{\begin{tikzpicture}
		\node[anchor=south west,inner sep=0] at (0,-0.15) {\includegraphics[width=.37\textwidth]{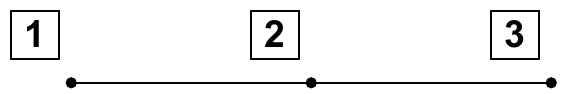}};
		\node at (2,0.3) {$q_1^{(0)}$};
		\node at (4.6,0.3) {$q_2^{(0)}$};
		\node at (7.5,0.3) {$q_3^{(0)}$};
		\draw[dashed,white] (-1.5,0) -- (0.6,0);
		\draw[dashed] (6.2,0) -- (8.2,0);
		\end{tikzpicture}}
	\subfigure{\begin{tikzpicture}
		\node[anchor=south west,inner sep=0] at (0,-0.15) {\includegraphics[width=.37\textwidth]{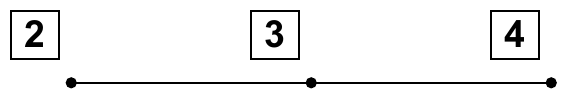}};
		\node at (2,0.3) {$q_2^{(0)}$};
		\node at (4.6,0.3) {$q_3^{(0)}$};
		\node at (-1,0.3) {$q_1^{(0)}$};
		\draw[dashed] (-1.5,0) -- (0.6,0);
		\draw[dashed,white] (6.4,0) -- (8.2,0);
		\end{tikzpicture}}
	\caption{Subgraphs associated to polynomials in Eqs.\eqref{poly-sing}. The $q$-function on the dashed edges correspond to nonremoved poles.}
\end{figure}

\begin{figure}[h!]
	\begin{center}
		\subfigure[]{\begin{tikzpicture}
			\node[anchor=south west,inner sep=0] at (0,-0.08) {\includegraphics[width=.2\textwidth]{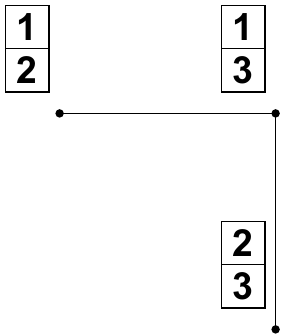}};
			\draw[dashed] (3.3,0) -- (5.5,0);
			\draw[dashed] (3.3,2.55) -- (5.5,2.55);
			\node at (1.8,2.9) {$q_2^{(1/2)}$};
			\node at (4.8,2.9) {$q_3^{(1/2)}$};
			\node at (4.8,.3) {$q_3^{(1/2)}$};
			\node at (3.9,1.6) {$q_1^{(-1/2)}$};
			\draw[dashed,white] (3.55,0) -- (3.55,-2.6);
			\draw[dashed,white] (6.3,0) -- (6.3,-2.6);
			\end{tikzpicture}\label{sl4-1223}}
		\hfill
		\subfigure[]{\begin{tikzpicture}
			\node[anchor=south west,inner sep=0] at (0,-0.15) {\includegraphics[width=.39\textwidth]{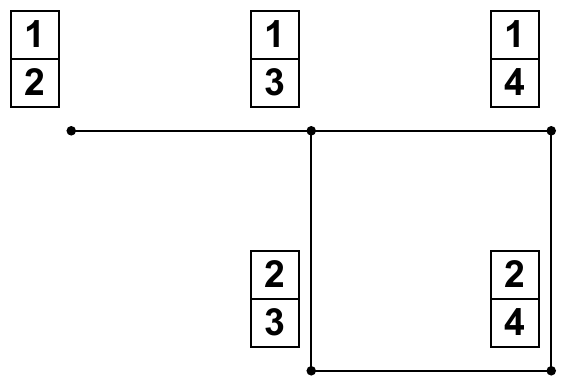}};
			\draw[dashed] (6.3,0) -- (6.3,-2.6);
			\node at (2,3.1) {$q_2^{(1/2)}$};
			\node at (4.9,3.1) {$q_3^{(1/2)}$};
			\node at (4.8,.3) {$q_3^{(1/2)}$};
			\node at (4.2,1.6) {$q_1^{(-1/2)}$};
			\node at (4.2,1.6) {$q_1^{(-1/2)}$};
			\node at (6.95,1.6) {$q_1^{(-1/2)}$};
			\node at (6.95,-1.3) {$q_2^{(-1/2)}$};
			\end{tikzpicture}
			\label{sl4-1224}}
		\vfill
		\subfigure[]{\begin{tikzpicture}
			\node[anchor=south west,inner sep=0] at (0,-0.4) {\includegraphics[width=.21\textwidth]{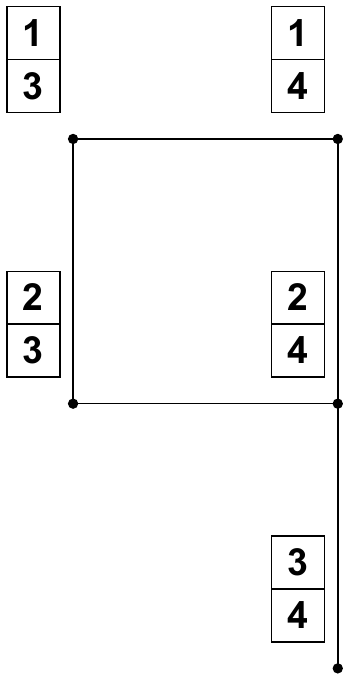}};
			\draw[dashed] (-1.8,5) -- (.6,5);
			\node at (1.9,2.7) {$q_3^{(1/2)}$};
			\node at (1.9,5.35) {$q_3^{(1/2)}$};
			\node at (-1.2,5.35) {$q_2^{(1/2)}$};
			\node at (1.25,3.95) {$q_1^{(1/2)}$};
			\node at (4,1.35) {$q_2^{(-1/2)}$};
			\node at (4,3.95) {$q_1^{(-1/2)}$};
			\end{tikzpicture}\label{sl4-1334}}
		\hfill
		\subfigure[]{\begin{tikzpicture}
			\node[anchor=south west,inner sep=0] at (0,-0.4) {\includegraphics[width=.2\textwidth]{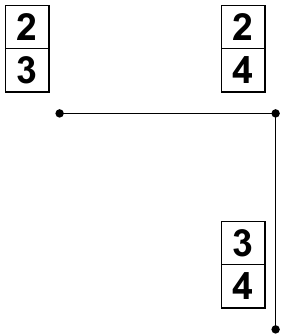}};
			\draw[dashed] (.7,2.35) -- (.7,5);
			\draw[dashed] (3.25,2.35) -- (3.25,5);
			\node at (1.8,2.6) {$q_3^{(1/2)}$};
			\node at (3.9,1.2) {$q_2^{(-1/2)}$};
			\node at (3.9,3.7) {$q_1^{(-1/2)}$};
			\node at (1.35,3.7) {$q_1^{(-1/2)}$};
			\end{tikzpicture}\label{sl4-2334}}
		\caption{Subgraphs associated to the polynomials defined in Eqs.\eqref{poly9}.}
		\label{sl4a2graphsub}
	\end{center}
\end{figure}
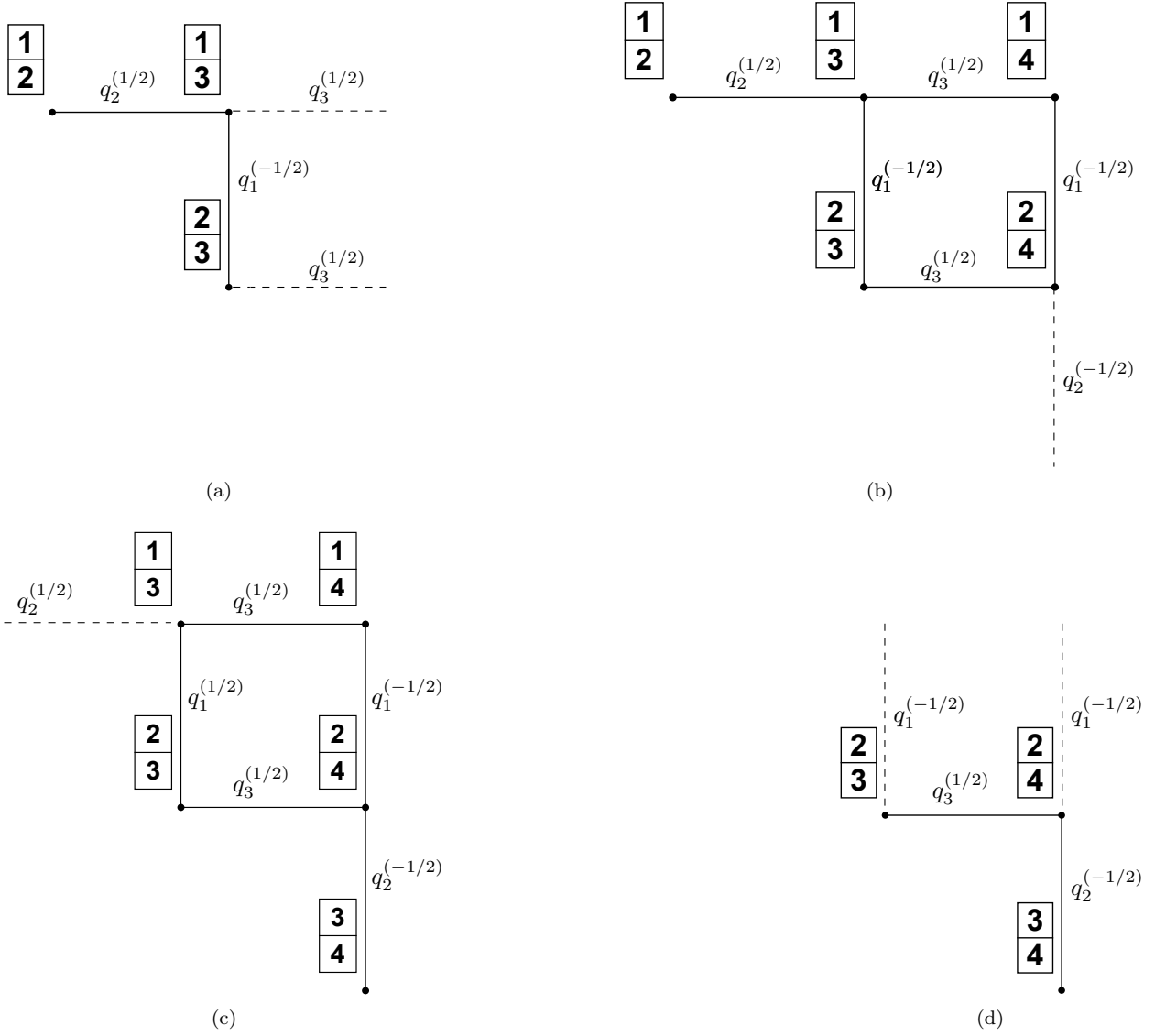

Now let us move on to the second fundamental representation. See Fig.\ref{sl4-2nd}. When handling larger graphs, the removal of a pole of a combination of terms of the eigenvalue expression requires one to take into account the several instances it may appear. For example, note that parallel edges like in $\begin{ytableau} 1
\\ 3 \end{ytableau} + \begin{ytableau} 2 \\ 3 \end{ytableau}$ and
$\begin{ytableau} 1 \\ 4 \end{ytableau} + \begin{ytableau} 2
\\ 4 \end{ytableau}$ remove the same pole at
$q_1(x-\frac{\im}{2})$. Therefore, any choice of three of these vertices does
not lead to a removal of this pole and one is forced to either stay with two of
them (non-diagonal) or all four. Therefore one cannot take just three vertices
of the center square. Moreover one must use the two dimensions of the graph to
obtain new {\unknown}s since straight lines would draw us back to $a=1$
graphs. Consequently all possible proper subgraphs are given in Fig.\ref{sl4a2graphsub}.

Note the center square alone is not included because it factorizes in terms of $a=1$ graphs:
\begin{equation}
\begin{ytableau} 1 \\ 3 \end{ytableau}+\begin{ytableau} 1 \\ 4 \end{ytableau} + \begin{ytableau} 2 \\ 3 \end{ytableau}   + \begin{ytableau} 2 \\ 4 \end{ytableau}
=\left(\begin{ytableau} 1 \\ \none[] \end{ytableau}+\begin{ytableau} 2 \\ \none \end{ytableau}\right) \times \left(\begin{ytableau} \none \\ 3 \end{ytableau}+\begin{ytableau} \none \\ 4 \end{ytableau}\right).
\end{equation}
Therefore, for the purpose of collecting the {\unknown}s it is desirable to avoid
factorization which breaks the graph into previously known parts. We also would like to highlight some special types of subgraphs that we call \textit{total}. These
are subgraphs where all filling indices $1,\ldots,n$ appear. For instance, only subgraphs in Fig.\ref{sl4-1224} and Fig.\ref{sl4-1334} are total. Total subgraphs have the property that they cannot appear in the lower rank models. Also, note that Fig.\ref{sl4-1223} corresponds to the $a=2$ $sl_3$ eigenvalue, while Fig.\ref{sl4-2334} is the same with the relabelling of fillings $j \rightarrow j+1$.

\section{Previous work on finite sets of NLIEs} \label{prev}
\ytableausetup{boxsize=normal}

In the following we write down the auxiliary functions obtained in previous works for $sl_n, n=2,3,4$.
\ytableausetup{mathmode,boxsize=2em,aligntableaux=center}

$sl_2$: {\color{blue} \cite{KluBat90,KluBatPea91,DesVeg92,KLUMPER92,KLUMPER93}}
	\begin{align}
	\mathsf{B}_1^{(2)}(x) &= \frac{\bt 1, 2 \et_x}{\bt 2 \et_x}, & 	\mathsf{B}_2^{(2)}(x) &= \frac{\bt 1, 2 \et_x}{\bt 1 \et_x}.
	\end{align}

$sl_3$: See \cite{KLUMPER-SU3}
\begin{subequations}
	\begin{align}
	\mathsf{B}_{1,1}^{(3)}(x) &= \frac{\bt 1 , 3 \et_{x+\im/2}}{\bt 2,3 \et_{x+\im/2}}, &
	\mathsf{B}_{1,2}^{(3)}(x) &= \frac{\bt 1 \\ 2,3 \et_x \bt  1,2 \\ 3 \et_x }{\bt 1 \\ 3 \et_x \bt 1,2 \\ 2,3 \et_x}, & 	\mathsf{B}_{1,3}^{(3)}(x) &= \frac{\bt 1,3 \et_{x-\im/2}}{\bt 1,2 \et_{x-\im/2}}, \\
 	\mathsf{B}_{2,1}^{(3)}(x) &= \frac{\bt 1,2 \\ 2,3 \et_{x+\im/2}}{\bt 1,2 \\ 3 \et_{x+\im/2}}, &
	\mathsf{B}_{2,2}^{(3)}(x) &= \frac{\bt 1,2 \et_x \bt 2,3 \et_x}{\bt 2 \et_x \bt 1,3 \et_x}, &
	\mathsf{B}_{2,3}^{(3)}(x) &= \frac{\bt 1,2 \\ 2,3 \et_{x-\im/2}}{\bt 1 \\ 2,3 \et_{x-\im/2}}.
	\end{align}
\end{subequations}

$sl_4$:  \cite{DAMERAU}
\begin{subequations}
	\begin{align}
	\mathsf{B}_{1,1}^{(4)}(x)&=\frac{\bt 1,4 \et_{x+\im/2}}{\bt 2,4 \et_{x+\im/2}}, &
	\mathsf{B}_{1,2}^{(4)}(x)&=\frac{\bt 1 \\ 2,4 \et_x \bt 1,3 \\ 3,4 \et_x}{\bt 1 \\ 3,4 \et_x \bt 1,3 \\ 2,4 \et_x}, &
	\mathsf{B}_{1,3}^{(4)}(x)&=\frac{\bt 1,3 \\ 4 \et_{x}\bt 1\\3,4 \et_{x}}{\bt 1 \\ 4 \et_x \bt 1,3 \\ 3,4 \et_x}, \label{jens-bb1} \\
	\mathsf{B}_{1,4}^{(4)}(x)&=\frac{\bt 1,4 \et_{x-\im/2}}{\bt 1,3 \et_{x-\im/2}}, &
	\mathsf{B}_{2,1}^{(4)}(x)&=\frac{\bt 1,3 \\ 2,4 \et_{x-\im/2}}{\bt 1,3 \\ 3,4 \et_{x-\im/2}}, &
	\mathsf{B}_{2,2}^{(4)}(x)&= \frac{\bt 2,3 \\ 4 \et_{x+\im/2} \bt 1,3 \\ 3,4\et_{x+\im/2}}{\bt 1,3 \\ 4 \et_{x+\im/2}\bt 2,3\\3,4 \et_{x+\im/2}}, \\
	\mathsf{B}_{2,3}^{(4)}(x)&=\frac{\bt 1,3 \et_x \bt 2,4 \et_x}{\bt 1,4 \et_x \bt 2,3 \et_x}, &
	\mathsf{B}_{2,4}^{(4)}(x)&=\frac{\bt 1 \\ 2,3 \\ 3,4 \et_x \bt 1,2 \\ 2,3 \\ 4 \et_x}{\bt 1 \\ 2,3 \\ 4 \et_x \bt 1,2 \\ 2,3 \\ 3,4 \et_x}, &
	\mathsf{B}_{2,5}^{(4)}(x)&= \frac{\bt 1 \\ 2,3 \et_{x-\im/2}\bt 1,2 \\ 2,4 \et_{x-\im/2}}{\bt 1\\ 2,4 \et_{x-\im/2}\bt 1,2 \\ 2,3 \et_{x-\im/2}}, \\
	\mathsf{B}_{2,6}^{(4)}(x)&=\frac{\bt 1,3 \\ 2,4 \et_{x-\im/2}}{\bt 1,2 \\ 2,4 \et_{x-\im/2}}, &
	\mathsf{B}_{3,1}^{(4)}(x)&=\frac{\bt 1,2 \\ 2,3 \\ 3,4 \et_{x+\im/2}}{\bt 1,2 \\ 2,3 \\ 4 \et_{x+\im/2}}, &
	\mathsf{B}_{3,2}^{(4)}(x)&=\frac{\bt 2 \\ 3,4 \et_{x} \bt 1,2 \\ 2,3 \et_x}{\bt 2\\3 \et_{x} \bt 1,2\\2,4 \et_x}, \\
	\mathsf{B}_{3,3}^{(4)}(x)&=\frac{\bt 2,3 \\ 3,4 \et_x \bt 1,2 \\ 2,4 \et_x}{\bt 2 \\ 3,4 \et_x \bt 1,3 \\ 2,4 \et_x},&
	\mathsf{B}_{3,4}^{(4)}(x)&=\frac{\bt 1,2 \\ 2,3 \\ 3,4 \et_{x-\im/2}}{\bt 1\\2,3\\3,4 \et_{x-\im/2}},\label{jens-bb14}
	\end{align}
\end{subequations}

\end{document}